\documentclass[aip,reprint,author-year,nofootinbib,floatfix]{revtex4-2}

\usepackage{graphicx}
\usepackage{amsmath,amssymb}             
\usepackage{stmaryrd}					
\usepackage[utf8]{inputenc}
\usepackage[T1]{fontenc}
\usepackage{color}
\usepackage[dvipsnames]{xcolor}
\usepackage{color}
\usepackage{todonotes}
\usepackage[raggedright]{titlesec}
\definecolor{darkblue}{rgb}{0.0,0.0,0.4}
\definecolor{red}{rgb}{0.7,0.0,0.0}
\definecolor{green}{rgb}{0.0,0.5,0.0}
\definecolor{C9}{HTML}{14BECF}

\usepackage{physjour_aas_macros}

\usepackage{tikz}
\usetikzlibrary{shapes,arrows,shadows}
\usepackage[export]{adjustbox}

\usepackage{scalerel}
\usetikzlibrary{svg.path}
\definecolor{orcidlogocol}{HTML}{A6CE39}
\tikzset{
  orcidlogo/.pic={
    \fill[orcidlogocol] svg{M256,128c0,70.7-57.3,128-128,128C57.3,256,0,198.7,0,128C0,57.3,57.3,0,128,0C198.7,0,256,57.3,256,128z};
    \fill[white] svg{M86.3,186.2H70.9V79.1h15.4v48.4V186.2z}
                 svg{M108.9,79.1h41.6c39.6,0,57,28.3,57,53.6c0,27.5-21.5,53.6-56.8,53.6h-41.8V79.1z M124.3,172.4h24.5c34.9,0,42.9-26.5,42.9-39.7c0-21.5-13.7-39.7-43.7-39.7h-23.7V172.4z}
                 svg{M88.7,56.8c0,5.5-4.5,10.1-10.1,10.1c-5.6,0-10.1-4.6-10.1-10.1c0-5.6,4.5-10.1,10.1-10.1C84.2,46.7,88.7,51.3,88.7,56.8z};
  }
}
\newcommand\orcid[1]{\href{https://orcid.org/#1}{\mbox{\scalerel*{
\begin{tikzpicture}[yscale=-1,transform shape]
\pic{orcidlogo};
\end{tikzpicture}
}{|}}}}

\usepackage{hyperref}
\hypersetup{colorlinks,breaklinks,
            linkcolor=darkblue,urlcolor=darkblue,
            anchorcolor=darkblue,citecolor=RoyalBlue}

\usepackage[mathscr]{euscript}
\usepackage{upgreek}
\DeclareFontFamily{OT1}{pzc}{}
\DeclareFontShape{OT1}{pzc}{m}{it}{<-> s * [1.10] pzcmi7t}{}
\DeclareMathAlphabet{\mathpzc}{OT1}{pzc}{m}{it}

\newcommand{\deriv}[2]{\dfrac{\mathrm{d} #1}{\mathrm{d} #2}}

\newcommand{\drm}{\mathrm{d}}

\newcommand{\G}{\mathrm{G}}

\titleformat{\section}{\selectfont \normalfont\raggedright\sffamily\small\bfseries\uppercase}{\thesection.}{1em}{}{}
\titleformat{\subsection}{\selectfont \normalfont\raggedright\sffamily\small\bfseries}{\thesubsection.}{1em}{}{}
\titleformat{\subsubsection}{\selectfont \normalfont\sffamily\small\bfseries}{\thesubsection.\thesubsubsection}{1em}{}{}
\titleformat{\paragraph}[runin]{\selectfont \normalfont\sffamily\small\bfseries}{\thesubsection.\thesubsubsection.\theparagraph}{1em}{}[.]

\begin{document}


\title{Perfectly parallel cosmological simulations using spatial comoving\newline Lagrangian acceleration}



\author{Florent~Leclercq}
\email{florent.leclercq@polytechnique.org}
\homepage{http://www.florent-leclercq.eu/}
\affiliation{Imperial Centre for Inference and Cosmology (ICIC) \& Astrophysics Group, Imperial College London, Blackett Laboratory, Prince Consort Road, London SW7 2AZ, United Kingdom}

\author{Baptiste~Faure}
\affiliation{AIM, CEA, CNRS, Université Paris-Saclay, Université Paris Diderot, Sorbonne Paris Cité, 91191 Gif-sur-Yvette, France}

\author{Guilhem~Lavaux}
\affiliation{CNRS, Sorbonne Université, UMR 7095, Institut d’Astrophysique de Paris, 98 bis bd Arago, 75014 Paris, France}

\author{Benjamin~D.~Wandelt}
\affiliation{CNRS, Sorbonne Université, UMR 7095, Institut d’Astrophysique de Paris, 98 bis bd Arago, 75014 Paris, France}
\affiliation{Sorbonne Universit\'e, Institut Lagrange de Paris (ILP), 98 bis bd Arago, 75014 Paris, France}
\affiliation{Center for Computational Astrophysics, Flatiron Institute, 162 5th Avenue, 10010, New York, NY, USA}

\author{Andrew~H.~Jaffe}
\affiliation{Imperial Centre for Inference and Cosmology (ICIC) \& Astrophysics Group, Imperial College London, Blackett Laboratory, Prince Consort Road, London SW7 2AZ, United Kingdom}

\author{Alan~F.~Heavens}
\affiliation{Imperial Centre for Inference and Cosmology (ICIC) \& Astrophysics Group, Imperial College London, Blackett Laboratory, Prince Consort Road, London SW7 2AZ, United Kingdom}

\author{Will~J.~Percival}
\affiliation{Waterloo Centre for Astrophysics, University of Waterloo, 200 University Ave W, Waterloo, ON N2L 3G1, Canada}
\affiliation{Department of Physics and Astronomy, University of Waterloo, 200 University Ave W, Waterloo, ON N2L 3G1, Canada}
\affiliation{Perimeter Institute for Theoretical Physics, 31 Caroline St. North, Waterloo, ON N2L 2Y5, Canada}

\author{Camille~Noûs}
\affiliation{Cogitamus Laboratory}


\date{\today}

\begin{abstract}
\noindent Existing cosmological simulation methods lack a high degree of parallelism due to the long-range nature of the gravitational force, which limits the size of simulations that can be run at high resolution. To solve this problem, we propose a new, perfectly parallel approach to simulate cosmic structure formation, which is based on the spatial COmoving Lagrangian Acceleration (sCOLA) framework. Building upon a hybrid analytical and numerical description of particles' trajectories, our algorithm allows for an efficient tiling of a cosmological volume, where the dynamics within each tile is computed independently. As a consequence, the degree of parallelism is equal to the number of tiles. We optimised the accuracy of sCOLA through the use of a buffer region around tiles and of appropriate Dirichlet boundary conditions around sCOLA boxes. As a result, we show that cosmological simulations at the degree of accuracy required for the analysis of the next generation of surveys can be run in drastically reduced wall-clock times and with very low memory requirements. The perfect scalability of our algorithm unlocks profoundly new possibilities for computing larger cosmological simulations at high resolution, taking advantage of a variety of hardware architectures.
\end{abstract}


\maketitle



\section{Introduction}

We live in the age of large astronomical surveys. These surveys detect and record tracers of cosmic structure across vast volumes of the Universe, using electromagnetic and gravitational waves. A non-exhaustive list includes optical and infrared imaging and spectroscopic surveys such as LSST \citep{LSSTScienceCollaboration2012}, Euclid \citep{Euclid}, DESI \citep{DESI}, and SPHEREx \citep{SPHEREx}; catalogues and intensity maps from large radio surveys such as the square kilometer array \citep{SKA} and its precursors; cluster catalogues from high-resolution observations of the microwave sky (Advanced ACTPol, \citealp{AdvACT}; SPTPol, \citealp{SPTPol}; Simons Observatory, \citealp{SimonsObservatory}, and CMB-S4); X-ray surveys such as the eROSITA mission \citep{eROSITA}; as well as gravitational wave sirens across cosmological volumes with successive updates of (Advanced) LIGO \citep{AdvancedLIGO}, Virgo \citep{AdvancedVirgo} and LISA \citep{LISA}. Whilst these data sets will be prodigious sources of scientific discovery across astrophysics, their enormous volume and dense sampling of cosmic structure will make them uniquely powerful when studying some of the deepest scientific mysteries of our time: the statistical properties of the primordial perturbations, the nature of dark matter, and the physical properties of dark energy. Indeed many of these surveys were conceived to address these questions.

Accomplishing this promise requires the ability to model these surveys in sufficient detail and with sufficient accuracy. All but the most simplistic models require the production of cosmological light-cone simulations. In particular, cosmological inferences often rely on large numbers of mock catalogues, which are used to construct unbiased estimators and study their statistical properties, such as covariance matrices. As surveys are getting deeper, these mock catalogues now need to represent a sizeable portion of the observable Universe, up to a redshift of $\sim 2-3$ (e.g. $z=2.3$ for the Euclid Flagship simulation\footnote{\url{https://www.euclid-ec.org/?page_id=4133}}). Unfortunately, cosmological simulations put a heavy load on supercomputers. Even if only dark matter is included and resolution is minimised, they can require millions of CPU hours and hundreds of terabytes of disk space to solve the gravitational evolution of billions of particles and store the corresponding data. For instance, the DEUS-FUR simulation \citep{DEUSFUR}, containing $8192^3$ particles in a box of $21~\mathrm{Gpc}/h$ side length, required $10$~million hours of CPU time and $300$~TB of storage. 

While computational needs are soaring, the performance of individual compute cores attained a plateau around 2015. Traditional hardware architectures are reaching their physical limit. Therefore, cosmological simulations cannot merely rely on processors becoming faster to reduce the computational time. Current hardware development focuses on increasing power efficiency\footnote{For example, Oak-Ridge National Laboratories' (ORNL) Summit machine has a typical power consumption of about $13$~MW.} and solving problems of heat dissipation to allow packing a larger number of cores into each CPU. As a consequence, the performance gains of the world's top supercomputers are the result of a massive increase in the number of parallel cores, currently\footnote{\url{https://www.olcf.ornl.gov/olcf-resources/compute-systems/summit/}} to $\mathcal{O}(10^5)$, and soon to $\mathcal{O}(10^{6-7})$ in systems that are currently being built.\footnote{See for example ORNL's next supercomputer, Frontier: \url{https://www.olcf.ornl.gov/wp-content/uploads/2019/05/frontier_specsheet.pdf}} Hybrid architectures, where CPUs work alongside GPUs and/or reconfigurable chips such as FPGAs, add to the massive parallelism. In the exa-scale world, raw compute cycles are no longer the scarce resource. The challenge is to access the available computational power when Amdahl's law demonstrates that communication latencies kill the potential gains due to parallelisation \citep{Amdahl1967}.

A way to embed high-resolution simulation of objects such as galaxy clusters, or even galaxies, in a cosmological context is through the use of varying particle mass resolution and the Adaptive Mesh Refinement technique \citep[AMR,][]{AMRpaper}. AMR is widely employed in grid-based simulation codes such as RAMSES \citep{RAMSES}, ENZO \citep{ENZO}, FLASH \citep{FLASH}, and AMIGA \citep{AMIGA}. It is also used in MUSIC \citep{Hahn2011} to generate zoom-in initial conditions for simulations. The AMR technique, which uses multi-grid relaxation methods \citep[e.g.][]{GuilletTeyssier2011}, allows focusing the effort on a specific region of the computational domain, but requires a two-way flow of information between small and large scales. More recently, leading computational cosmology groups have been developing sophisticated schemes to leverage parallel and hybrid computing architectures \citep{Gonnet2013,Theuns2015,Aubert2015,Ocvirk2016,Potter2017,Yu2018,Garrison2019,Cheng2020}.

Full simulations of large cosmological volumes, even limited to cold dark matter and at coarse resolution, involve multiple challenges. One of the main issues preventing their easy parallelisation is the long-range nature of gravitational interactions, which forestalls high-resolution, large-volume cosmological simulations. As a a response, much of the classical work in numerical cosmology focused on computational algorithms (tree codes, fast multipole methods, particle-mesh methods, and hybrids such as particle-particle--particle-mesh and tree--particle-mesh) that reduced the need for $\mathcal{O}(N^2)$ all-to-all communications between $N$ particles across the full computational volume.

While these algorithms are and remain the backbone of computational cosmology, they fail to fully exploit the physical scale hierarchy of cosmological perturbations. This hierarchy has first been used to push the results of $N$-body simulations to Universe scale for cosmic velocity fields \citep{Strauss1995}. At the largest scales, the dynamics of the Universe is not complicated, and in particular, is well-captured by Lagrangian Perturbation Theory \citep[LPT; see][]{Bouchet1995}. Building upon this view, \citet{Tassev2015} introduced spatial COmoving Lagrangian Acceleration (sCOLA). This algorithm, using a hybrid analytical and numerical treatment of particles' trajectories, allows one to perform simulations without the need to substantially extend the simulated volume beyond the region of interest in order to capture far-field effects, such as density fluctuations due to super-box modes. The sCOLA proof-of-concept focused on one sub-box embedded into a larger simulation box.

In this paper, we extend the sCOLA algorithm and use it within a novel method for perfectly parallel cosmological simulations. To do so, we rely on a tiling of the full cosmological volume to be simulated, where each tile is evolved independently using sCOLA. The principal challenge for the accuracy of such simulations are the boundary conditions used throughout the evolution of tiles, which can introduce artefacts. In this respect, we introduce three crucial improvements with respect to \citet{Tassev2015}: the use of a buffer region around each tile, the use of exact boundary conditions in the calculation of LPT displacements (which has the side benefit of reducing memory requirements), and the use of a Poisson solver with Dirichlet boundary conditions meant to approximate the exact gravitational potential around sCOLA boxes. The method proposed in this work shares similar goals with zoom-in simulation techniques, the main difference residing in the change of frame of reference introduced in sCOLA, which accounts for the dynamics of large scales without requiring flows of information during the evolution. On the other hand, our method is independent of the $N$-body integrator used to calculate the numerical part of particles' trajectories within each sCOLA box, and therefore, it cannot be related to specific approaches to do so, such as force-splitting. It is slightly approximate and more CPU-expensive than the corresponding ``monolithic'' simulation technique \citep[chosen in this paper as tCOLA,][]{Tassev2013}, but has the essential advantage of perfect scalability. This scalability comes from the removal of any kind of communication among tiles after the initialisation of the simulation. As a consequence, for its major part, the degree of parallelism of the algorithm equals the number of tiles, which means that the workload is perfectly parallel (also called embarrassingly parallel). This property can be exploited to produce cosmological simulations in very short wall-clock times on a variety of hardware architectures, as we discuss in this paper.

After reviewing Lagrangian Perturbation Theory and its use within numerical simulations in section \ref{sec:Cosmological simulations using Lagrangian Perturbation Theory}, we describe our algorithm for perfectly parallel cosmological simulations in section \ref{sec:Algorithm for perfectly parallel simulations using sCOLA}. In section \ref{sec:Accuracy and speed}, we test the accuracy and speed of the algorithm with respect to reference simulations that do not use the tiling. We discuss the implications of our results for computational strategies to model cosmic structure formation, and conclude, in section \ref{sec:Conclusion}. Details regarding the implementation are provided in the appendices.

\section{Cosmological simulations using Lagrangian perturbation theory}
\label{sec:Cosmological simulations using Lagrangian Perturbation Theory}

Throughout this section we denote by $a$ the scale factor of the Universe. For simplicity, some of the equations are abridged. We reintroduce the omitted constants, temporal prefactors, and Hubble expansion in appendix \ref{apx:Model equations}.

Particle simulators are algorithms that compute the final position $\textbf{x}$ and momentum $\textbf{p} \equiv \drm \textbf{x}/ \drm a$ of a set of particles, given some initial conditions. They can also be seen as algorithms that compute a displacement field $\boldsymbol{\Psi}$, which maps the initial (Lagrangian) position $\textbf{q}$ of each particle to its final (Eulerian) position $\textbf{x}$, according to the classic equation \citep[see e.g.][for a review]{Bernardeau2002}
\begin{equation}
\textbf{x}(a) = \textbf{q} + \boldsymbol{\Psi}(\textbf{q},a) .
\end{equation}
With this point of view, the outputs are $\textbf{x}$ and $\textbf{p}~=~\partial \boldsymbol{\Psi} / \partial a$.

\subsection{Lagrangian perturbation theory (LPT)}
\label{ssec:Lagrangian perturbation theory (LPT)}

In Lagrangian perturbation theory (LPT), the displacement field is given by an analytic equation which is used to move particles, without the need for a numerical solver. At second order in LPT, the displacement field is written
\begin{equation}
\boldsymbol{\Psi}_\mathrm{LPT}(\textbf{q},a) = \boldsymbol{\Psi}^{(1)}(\textbf{q},a) + \boldsymbol{\Psi}^{(2)}(\textbf{q},a),
\label{eq:Psi_LPT}
\end{equation}
where each of the terms is separable into a temporal and a spatial contribution deriving from a Lagrangian potential:
\begin{eqnarray}
\boldsymbol{\Psi}^{(1)}(\textbf{q},a) & = & -D_1(a) \, \boldsymbol{\nabla}_\textbf{q} \phi^{(1)}(\textbf{q}),\label{eq:LPT_Psi1}\\
\boldsymbol{\Psi}^{(2)}(\textbf{q},a) & = & D_2(a) \, \boldsymbol{\nabla}_\textbf{q} \phi^{(2)}(\textbf{q})\label{eq:LPT_Psi2}.
\end{eqnarray}

In equations \eqref{eq:LPT_Psi1} and \eqref{eq:LPT_Psi2}, $D_1$ and $D_2$ are the growth factor and second-order growth factor, respectively. The Lagrangian potentials obey Poisson-like equations \citep{Buchert1994}:
\begin{eqnarray}
\boldsymbol{\Delta}_\textbf{q} \phi^{(1)}(\textbf{q}) & = & \delta_\mathrm{i}(\textbf{q}), \label{eq:LPTpotential1}\\
\boldsymbol{\Delta}_\textbf{q} \phi^{(2)}(\textbf{q}) & = & \sum_{i>j} \left[ \phi^{(1)}_{,ii}\phi^{(1)}_{jj} - \left(\phi^{(1)}_{,ij}\right)^2 \right], \label{eq:LPTpotential2}
\end{eqnarray}
where $\delta_\mathrm{i}(\textbf{q})$ is the density contrast in the initial conditions, in Lagrangian coordinates, and the $\phi^{(1)}_{,ij}$ are spatial second derivatives of $\phi^{(1)}$, i.e. $\phi^{(1)}_{,ij} \equiv \partial^2 \phi^{(1)}/\partial \textbf{q}_i \partial \textbf{q}_j$.

If only the first-order term is included in equation \eqref{eq:Psi_LPT}, the solution is known as the Zel'dovich approximation \citep{Zeldovich1970}.

\subsection{Temporal comoving Lagrangian acceleration (tCOLA)}
\label{csec:Temporal comoving Lagrangian acceleration (tCOLA)}

In contrast to the analytical equations of LPT, particle-mesh (PM) codes \citep[see e.g.][]{Klypin1997} provide a fully numerical solution to the problem of large-scale structure formation. The equation of motion to be solved in a PM code reads schematically 
\begin{equation}
\partial_a^2 \boldsymbol{\Psi}(\textbf{q},a) = -\boldsymbol{\nabla}_\textbf{x} \Phi(\textbf{x},a),
\label{eq:PM_EoM}
\end{equation}
where the gravitational potential $\Phi$ satisfies the Poisson equation,
\begin{equation}
\Delta_\textbf{x} \Phi(\textbf{x},a) = \delta(\textbf{x},a).
\label{eq:Poisson_full_box}
\end{equation}
Here, $\delta(\textbf{x},a)$ is the density contrast at a scale factor $a$, which is obtained from the set of particles' positions $\left\lbrace \textbf{x}(a) \right\rbrace$ through a density assignment operator that we denote $\mathrm{B}$ \citep[typically a cloud-in-cell (CiC) scheme, see][]{Hockney1981}:
\begin{equation}
\delta(\textbf{x},a) \equiv \mathrm{B}(\left\lbrace \textbf{x}(a) \right\rbrace).
\end{equation}
We denote by $\bar{\mathrm{B}}$ the corresponding interpolation operator, which is needed to obtain the accelerations of particles given the acceleration field on the grid:
\begin{equation}
\partial_a^2 \boldsymbol{\Psi}(\left\lbrace \textbf{x}(a) \right\rbrace) \equiv \bar{\mathrm{B}}(-\boldsymbol{\nabla}_\textbf{x} \Phi).
\end{equation}

The temporal COmoving Lagrangian Acceleration (tCOLA) algorithm seeks to decouple large and small scales by evolving large scales using analytic LPT results, and small scales using a numerical solver. This is achieved by splitting the Lagrangian displacement field into two contributions \citep{Tassev2012b}:
\begin{equation}
\boldsymbol{\Psi}(\textbf{q},a) \equiv \boldsymbol{\Psi}_\mathrm{LPT}(\textbf{q},a) + \boldsymbol{\Psi}_\mathrm{res}(\textbf{q},a),
\label{eq:displacement_split}
\end{equation}
where $\boldsymbol{\Psi}_\mathrm{LPT}(\textbf{q},a)$ is the LPT displacement field discussed in section \ref{ssec:Lagrangian perturbation theory (LPT)} and $\boldsymbol{\Psi}_\mathrm{res}(\textbf{q},a)$ is the residual displacement of each particle, as measured in a frame comoving with an ``LPT observer'', whose trajectory is given by $\boldsymbol{\Psi}_\mathrm{LPT}(\textbf{q},a)$. Using equation \eqref{eq:displacement_split}, it is possible to rewrite equation \eqref{eq:PM_EoM} as
\begin{equation}
\partial_a^2 \boldsymbol{\Psi}_\mathrm{res}(\textbf{q},a) = -\boldsymbol{\nabla}_\textbf{x} \Phi(\textbf{x},a) - \partial_a^2 \boldsymbol{\Psi}_\mathrm{LPT}(\textbf{q},a).
\label{eq:tCOLA_EoM}
\end{equation}
The term $\partial_a^2 \boldsymbol{\Psi}_\mathrm{LPT}(\textbf{q},a)$ can be thought of as a fictitious force acting on particles, caused by our use of a non-inertial frame of reference. Importantly, it can be computed analytically given the equations of Lagrangian perturbation theory.

The equations of motions \eqref{eq:PM_EoM} and \eqref{eq:tCOLA_EoM} are usually integrated by the use of time-stepping techniques (see appendix \ref{apx:Standard and modified time-stepping}). In the limit of zero time-steps used to discretise the left-hand side of equation \eqref{eq:tCOLA_EoM}, $\boldsymbol{\Psi}_\mathrm{res}=0$ and tCOLA recovers the results of LPT; therefore, tCOLA always solves the large scales with an accuracy of at least that of LPT. In contrast, PM codes require many time-steps in equation \eqref{eq:PM_EoM} just to recover the value of the linear growth factor $D_1$. In the limit where the number of time-steps becomes large, tCOLA reduces to a standard PM code. In the intermediate regime (for $\mathcal{O}(10)$ time-steps), tCOLA provides a good approximation to large-scale structure formation, at the expense of not solving the details of particle trajectories in deeply non-linear halos \citep[see][for further discussion]{Tassev2013,Howlett2015,Leclercq2015ST,Koda2016,Izard2016}. Since by construction, tCOLA always gets the large scales correct, contrary to a PM code, the trade-off between speed and accuracy only affects small scales.

\subsection{Spatial comoving Lagrangian acceleration (sCOLA)}
\label{ssec:Spatial comoving Lagrangian acceleration (sCOLA)}

During large-scale structure formation, non-linearities appear at late times and/or at small scales. tCOLA (equation \eqref{eq:tCOLA_EoM}) decouples LPT displacements and residual non-linear contributions ``in time'', so that, for a given accuracy, fewer time-steps are required to solve large-scale structure evolution than with a PM code. Following a similar spirit, the spatial COmoving Lagrangian Acceleration (sCOLA) framework decouples LPT displacements and residual non-linear contributions ``in space'', so that numerically evolved small scales can feel far-field effects captured analytically via LPT.

More specifically, for each particle in a volume of interest (the ``sCOLA box'') embedded in a larger cosmological volume (the ``full box''), the equation of motion of particles, which reads for a traditional $N$-body problem
\begin{equation}
\partial_a^2 \boldsymbol{\Psi}(\textbf{q},a) = \partial_a^2 \boldsymbol{\Psi}_\mathrm{LPT}(\textbf{q},a) + \partial_a^2 \boldsymbol{\Psi}_\mathrm{res}(\textbf{q},a) = \textbf{F}(\textbf{x},a)
\label{eq:NbodyEoM}
\end{equation}
is replaced by
\begin{equation}
\partial_a^2 \boldsymbol{\Psi}_\mathrm{res}(\textbf{q},a) = \textbf{F}^\mathrm{sCOLA}(\textbf{x},a) - \partial_a^2 \boldsymbol{\Psi}^\mathrm{sCOLA}_\mathrm{LPT}(\textbf{q},a) .
\label{eq:sCOLA_NbodyEoM}
\end{equation}
$\partial_a^2 \boldsymbol{\Psi}_\mathrm{res}(\textbf{q},a)$ is defined by equation \eqref{eq:displacement_split} as the residual displacement with respect to the LPT observer of the full box, whose trajectory is given by $\boldsymbol{\Psi}_\mathrm{LPT}(\textbf{q},a)$. In equation \eqref{eq:sCOLA_NbodyEoM}, $\boldsymbol{\Psi}_\mathrm{LPT}^\mathrm{sCOLA}(\textbf{q},a)$ is the trajectory prescribed by solving LPT equations (see section \ref{ssec:Lagrangian perturbation theory (LPT)}) in the sCOLA box. Note that $\boldsymbol{\Psi}_\mathrm{LPT}^\mathrm{sCOLA}(\textbf{q},a)$ may differ from $\boldsymbol{\Psi}_\mathrm{LPT}(\textbf{q},a)$, depending on the assumptions made for the boundary conditions of the sCOLA box, discussed in section \ref{ssec:Initial operations in the sCOLA boxes}. Denoting by $\mathcal{S} \subseteq \llbracket 1,N \rrbracket$ the set of particles in the sCOLA box, the gravitational force, which in equation \eqref{eq:NbodyEoM} reads
\begin{equation}
\textbf{F}(\textbf{x}_i,a) \equiv \sum_{j=1 \atop j \neq i}^N \frac{\textbf{x}_j(a)-\textbf{x}_i(a)}{|\textbf{x}_j(a)-\textbf{x}_i(a)|^3},
\label{eq:force_full}
\end{equation}
is replaced by
\begin{equation}
\textbf{F}^\mathrm{sCOLA}(\textbf{x}_i,a) \equiv \sum_{j\in \mathcal{S} \atop j \neq i} \frac{\textbf{x}_j(a)-\textbf{x}_i(a)}{|\textbf{x}_j(a)-\textbf{x}_i(a)|^3}.
\label{eq:force_sCOLA}
\end{equation}

It is possible to evaluate $\textbf{F}^\mathrm{sCOLA}(\textbf{x},a)$, and thus to solve equation \eqref{eq:sCOLA_NbodyEoM}, like equation \eqref{eq:NbodyEoM}, using any numerical gravity solver, such as particle-particle--particle-mesh, tree codes, or AMR. In this paper, we choose to focus on evaluating forces via a PM scheme. In this case,  the equation of motion of particles in sCOLA reads schematically \citep{Tassev2015}
\begin{equation}
\partial_a^2 \boldsymbol{\Psi}_\mathrm{res}(\textbf{q},a) = -\boldsymbol{\nabla}_\textbf{x}^\mathrm{sCOLA} \Phi^\mathrm{sCOLA}(\textbf{x},a) - \partial_a^2 \boldsymbol{\Psi}_\mathrm{LPT}^\mathrm{sCOLA}(\textbf{q},a) .
\label{eq:sCOLA_EoM}
\end{equation}
The gravitational potential in the sCOLA box, $\Phi^\mathrm{sCOLA}(\textbf{x},a)$, obeys the near-field version of the Poisson equation,
\begin{equation}
\Delta_\textbf{x}^\mathrm{sCOLA} \Phi^\mathrm{sCOLA}(\textbf{x},a) = \delta^\mathrm{sCOLA}(\textbf{x},a) .
\label{eq:Poisson_sCOLA_box}
\end{equation}
The superscript ``sCOLA'' over the gradient and Laplacian operators, $\boldsymbol{\nabla}_\textbf{x}^\mathrm{sCOLA}$ and $\Delta_\textbf{x}^\mathrm{sCOLA}$, mean that they are restricted to the sCOLA box (contrary to that of equations \eqref{eq:Poisson_full_box} and \eqref{eq:tCOLA_EoM}). Over the density contrast $\delta^\mathrm{sCOLA}(\textbf{x},a)$, the superscript means that only particles in the sCOLA box $\left\lbrace \textbf{x}(a) \right\rbrace_\mathrm{sCOLA} \equiv \left\lbrace \textbf{x}_i(a) \right\rbrace_{i\in \mathcal{S}}$ (instead of the full box) are used within the density assignment $\mathrm{B}^\mathrm{sCOLA}$, i.e.
\begin{equation}
\delta^\mathrm{sCOLA}(\textbf{x},a) \equiv \mathrm{B}^\mathrm{sCOLA}\left(\left\lbrace \textbf{x}(a) \right\rbrace_\mathrm{sCOLA}\right).
\end{equation}

Contrary to tCOLA, which is an exact rewriting of the equations of motion of a PM code, sCOLA potentially involves approximations for the calculation of each quantity and operator with a superscript ``sCOLA'' instead of its full box equivalent. As a proof of concept, \citet{Tassev2015} showed that under certain circumstances, sCOLA provides a good approximation for the evolution of one sCOLA box embedded into a larger full box. As discussed in the introduction, we aim at generalising this result by using sCOLA within multiple sub-volumes of a full simulation box.

\section{Algorithm for perfectly parallel simulations using sCOLA}
\label{sec:Algorithm for perfectly parallel simulations using sCOLA}

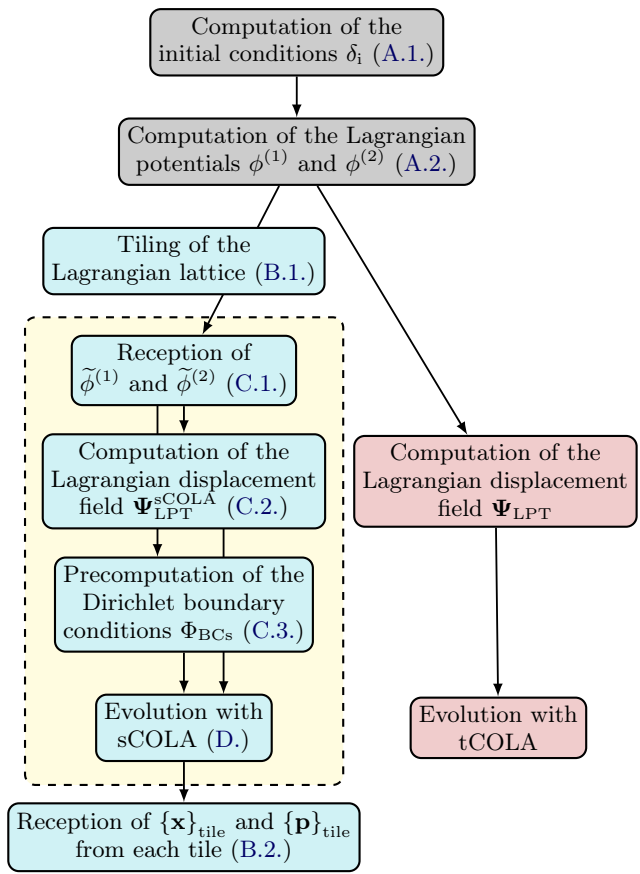
\begin{figure}
\begin{tikzpicture}
	\pgfdeclarelayer{background}
	\pgfdeclarelayer{foreground}
	\pgfsetlayers{background,main,foreground}

	\tikzstyle{common}=[draw, align=center, thick, rounded corners, minimum height=1em, minimum width=1em, fill=black!20]
	\tikzstyle{sCOLA}=[draw, align=center, thick, rounded corners, minimum height=1em, minimum width=1em, fill=C9!20]
	\tikzstyle{tCOLA}=[draw, align=center, thick, rounded corners, minimum height=1em, minimum width=1em, fill=red!20]
	\tikzstyle{plate} = [draw, thick, rectangle, rounded corners, dashed, fill=yellow!15]

	\def\blockdist{1.0}
	\def\modeldist{1.5}

    \node (ICs) [common]
    {Computation of the\\ initial conditions $\delta_\mathrm{i}$ (\hyperref[ssec:Initial conditions and Lagrangian potentials]{A.1.})};
    \path (ICs.south)+(0,-\blockdist) node (phifullbox) [common]
    {Computation of the Lagrangian\\ potentials $\phi^{(1)}$ and $\phi^{(2)}$ (\hyperref[ssec:Initial conditions and Lagrangian potentials]{A.2.})};
    \path (phifullbox.south)+(-\modeldist,-\blockdist) node (tiling) [sCOLA]
    {Tiling of the\\ Lagrangian lattice (\hyperref[ssec:Tiling and buffer region]{B.1.})};
    \path (tiling.south)+(0,-\blockdist) node (phisCOLAbox) [sCOLA]
    {Reception of\\ $\widetilde{\phi}^{(1)}$ and $\widetilde{\phi}^{(2)}$ (\hyperref[ssec:Initial operations in the sCOLA boxes]{C.1.})};
    \path (phisCOLAbox.south)+(0,-\blockdist) node (LPTsCOLA) [sCOLA]
    {Computation of the\\ Lagrangian displacement\\ field $\boldsymbol{\Psi}_\mathrm{LPT}^\mathrm{sCOLA}$ (\hyperref[ssec:Initial operations in the sCOLA boxes]{C.2.})};
    \path (LPTsCOLA.east)+(1.5*\modeldist,0) node (LPTtCOLA) [tCOLA]
    {Computation of the\\ Lagrangian displacement\\ field $\boldsymbol{\Psi}_\mathrm{LPT}$};
    \path (LPTsCOLA.south)+(0,-\blockdist) node (phiBCs) [sCOLA]
    {Precomputation of the\\ Dirichlet boundary\\ conditions $\Phi_\mathrm{BCs}$ (\hyperref[ssec:Initial operations in the sCOLA boxes]{C.3.})};
    \path (phiBCs.south)+(0,-\blockdist) node (sCOLAevolution) [sCOLA]
    {Evolution with\\ sCOLA (\hyperref[ssec:Evolution of sCOLA boxes]{D.})};
    \path (sCOLAevolution.east)+(2*\modeldist,0) node (tCOLAevolution) [tCOLA]
    {Evolution with\\ tCOLA};
    \path (sCOLAevolution.south)+(0,-\blockdist) node (untiling) [sCOLA]
    {Reception of $\left\lbrace\textbf{x}\right\rbrace_\mathrm{tile}$ and $\left\lbrace\textbf{p}\right\rbrace_\mathrm{tile}$\\ from each tile (\hyperref[ssec:Tiling and buffer region]{B.2.})};

    \begin{pgfonlayer}{background}    
    \path [plate] (tiling.south)+(-6.5em,-0.9em) rectangle (1.9em,-30.4em)
    {};
	\path [draw, line width=0.7pt, arrows={-latex}] (ICs) -- (phifullbox);
	\path [draw, line width=0.7pt, arrows={-latex}] (phifullbox) -- (LPTtCOLA);
	\path [draw, line width=0.7pt, arrows={-latex}] (LPTtCOLA) -- (tCOLAevolution);
	\path [draw, line width=0.7pt, arrows={-latex}] (phisCOLAbox) -- (LPTsCOLA);
	\path [draw, line width=0.7pt, arrows={-latex}] (phiBCs) -- (sCOLAevolution);
	\path [draw, line width=0.7pt, arrows={-latex}] (sCOLAevolution) -- (untiling);
	\path [draw, line width=0.7pt, arrows={-latex}] (phifullbox) -- (phisCOLAbox);
	\path (phisCOLAbox.south)+(-10pt,3.5pt) node (phisCOLAboxS) {};
	\path (phiBCs.north)+(-10pt,-3.5pt) node (phiBCsN) {};
	\path [draw, line width=0.7pt, arrows={-latex}] (phisCOLAboxS) -- (phiBCsN);
	\path (LPTsCOLA.south)+(15pt,3.5pt) node (LPTsCOLAS) {};
	\path (sCOLAevolution.north)+(15pt,-3.5pt) node (sCOLAevolutionN) {};
	\path [draw, line width=0.7pt, arrows={-latex}] (LPTsCOLAS) -- (sCOLAevolutionN);
    \end{pgfonlayer}

\end{tikzpicture}
\caption{Functional diagram of sCOLA (left) versus tCOLA (right). The grey boxes are common steps. sCOLA specific steps are represented in blue, and tCOLA specific steps in red. The yellow rectangle constitutes the perfectly parallel section, within which no communication is required with the master process or between processes. Arrows represent dependencies, and references to the main text are given between parentheses.}
\label{fig:sCOLA_diagram}
\end{figure}

\begin{table}[]
    \centering
    \begin{tabular}{cl}
         \hline
         \hline
         Symbol & Meaning \\
         \hline
         $N$ & \footnotesize{LPT grid size in the full box} \\
         $N_\mathrm{p}$ & \footnotesize{Lagrangian lattice size in the full box} \\
         $N_\mathrm{tiles}$ & \footnotesize{Number of tiles in each direction} \\
         $N_\mathrm{p,tile}$ & \footnotesize{Number of particles per direction in each tile} \\
         $L_\mathrm{tile}$ & \footnotesize{Physical size of each tile} \\
		 $N_\mathrm{p,buffer}$ & \footnotesize{Number of buffer particles per direction} \\
         $L_\mathrm{buffer}$ & P\footnotesize{hysical size of the buffer region} \\
         $N_\mathrm{p,sCOLA}$ & \footnotesize{Number of particles per direction in each sCOLA box} \\
         $L_\mathrm{sCOLA}$ & \footnotesize{Physical size of each sCOLA box} \\
         $N_\mathrm{tile}$ & \footnotesize{LPT grid portion covering each tile} \\
         $N_\mathrm{sCOLA}$ & \footnotesize{LPT grid portion covering each sCOLA box} \\
         $N_\mathrm{ghost}$ & \footnotesize{Number of ghost cells depending on FDA} \\
         $N_\mathrm{g}$ & \footnotesize{PM grid size in each sCOLA box} \\
         $r$ & \footnotesize{Over-simulation factor} \\
         $p$ & \footnotesize{Parallelisation potential factor} \\
		\hline\hline
    \end{tabular}
    \caption{Nomenclature of symbols used in the present article.}
    \label{tb:nomenclature}
\end{table}

\begin{figure}
\includegraphics[width=\linewidth]{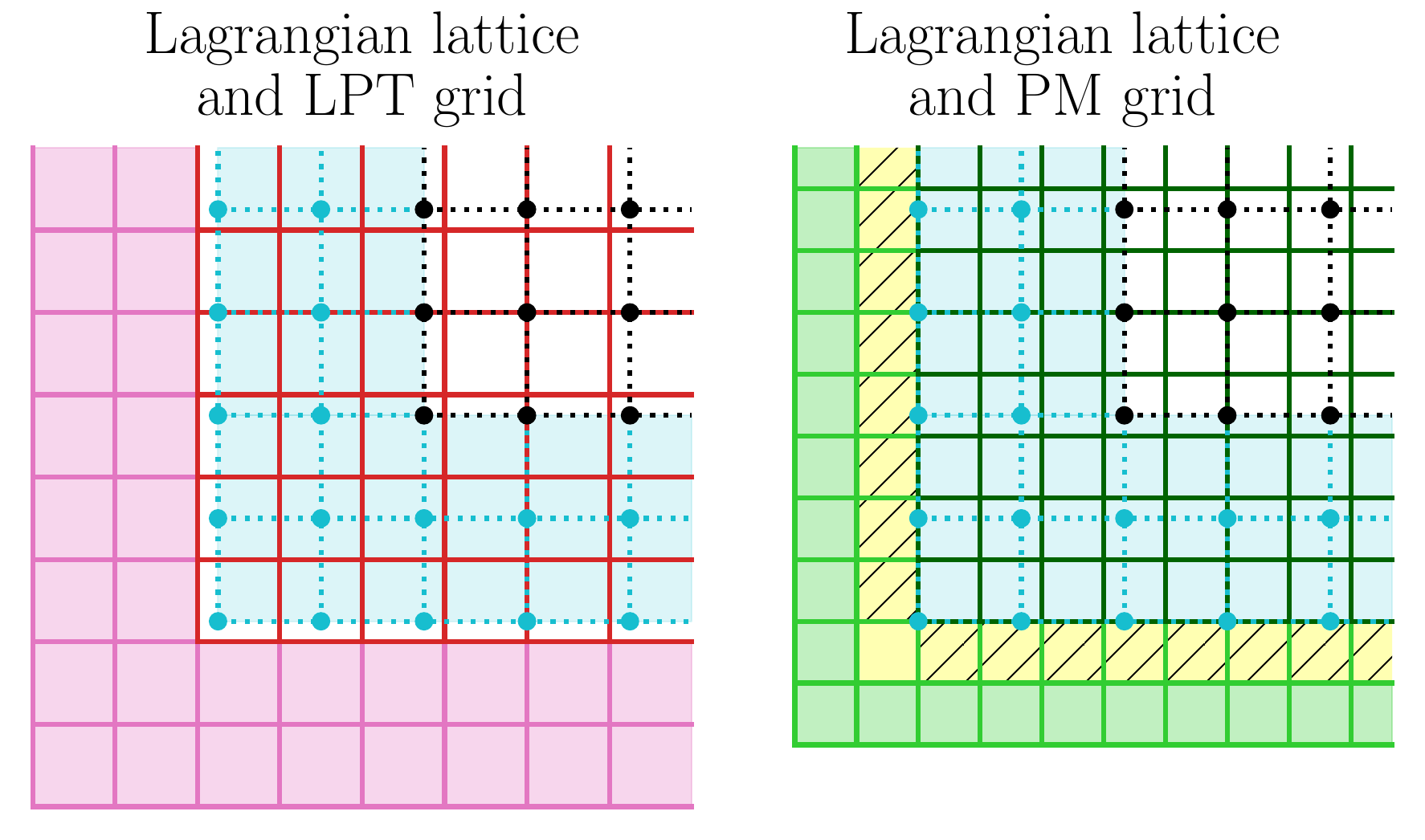} 
\caption{Illustration of the different grids used within sCOLA. The Lagrangian lattice is represented by dashed lines. For each tile, central particles (in black) are surrounded by buffer particles (in cyan), which are ignored at the end of the evolution. The corresponding buffer region in other grids is represented in cyan. The left panel represents the ``LPT grid'' on which Lagrangian potentials $\widetilde{\phi}^{(1)}$ and $\widetilde{\phi}^{(2)}$ are defined. The central region has $N_\mathrm{sCOLA}^3$ grid points (in red) and is padded by $2N_\mathrm{ghost}$ cells in each direction (pink region). The right panel shows the ``PM grid'' on which the density contrast $\delta^\mathrm{sCOLA}$, the gravitational potential $\Phi^\mathrm{sCOLA}$, and the accelerations $-\boldsymbol{\nabla}_\textbf{x}^\mathrm{sCOLA} \Phi^\mathrm{sCOLA}$ are defined. The density contrast is defined only in the central region (which has $N_\mathrm{g}^3$ grid points, in dark green). The gravitational potential is padded by $2N_\mathrm{ghost}$ cells in each direction (light green and yellow regions), and the gridded accelerations only by $N_\mathrm{ghost}$ cells in each direction (yellow region). Solving the Poisson equation requires Dirichlet boundary conditions in six layers of $N_\mathrm{ghost}$ cells, denoted as hatched regions. For simplicity of representation, we have used here $N_\mathrm{ghost}=1$.}
\label{fig:sCOLA_grids}
\end{figure}

In this section, we describe an algorithm for cosmological simulations using sCOLA, for which the time evolution of independent Lagrangian sub-volumes is perfectly parallel, without any communication. A functional block diagram representing the main steps and their dependencies is given in figure \ref{fig:sCOLA_diagram}. An illustration of the different grids appearing in the algorithm is presented in figure \ref{fig:sCOLA_grids}, and table \ref{tb:nomenclature} provides the nomenclature of some of the different variables appearing in this section.

We work in a cubic full box of side length $L$ with periodic boundary conditions, populated by $N_\mathrm{p}^3$ particles initially at the nodes $\left\lbrace \textbf{q} \right\rbrace$ of a regular Lagrangian lattice. We seek to compute the set of final positions $\left\lbrace \textbf{x}(a_\mathrm{f}) \right\rbrace$ and momenta $\left\lbrace \textbf{p}(a_\mathrm{f}) \right\rbrace$ at final scale factor $a_\mathrm{f}$. The model equations are reviewed in appendix \ref{apx:Model equations}. The time-stepping of these equations consists of a series of ``kick'' and ``drift'' operations and is discussed in appendix \ref{apx:Standard and modified time-stepping}.

We approximate the Laplacians $\Delta_\textbf{x}$, $\Delta_\textbf{q}$ and gradient operators $\boldsymbol{\nabla}_\textbf{x}$, $\boldsymbol{\nabla}_\textbf{q}$ by finite difference approximation (FDA) at order 2, 4, or 6. The coefficients of the finite difference stencils in configuration and in Fourier space are given for example in table 1 in \citet{Hahn2011}. We note $N_\mathrm{ghost} = 1, 2, 3$ if FDA is taken at order 2, 4, 6, respectively.

\subsection{Initial conditions and Lagrangian potentials}
\label{ssec:Initial conditions and Lagrangian potentials}

Before the perfectly parallel section, two initialisation steps are performed by the master process in the full box.
\begin{enumerate}
\item[A.1.] The first step is to generate the initial density contrast $\delta_\mathrm{i}$ in the full box, on a cubic grid of $N^3$ cells (the ``LPT grid'', represented in red in the left panel of figure \ref{fig:sCOLA_grids}). This step can be done via the standard convolution approach \citep[e.g.][]{Hockney1981}, given the specified initial power spectrum.
\item[A.2.] The second step is to compute the Lagrangian potentials $\phi^{(1)}(\textbf{q})$ and $\phi^{(2)}(\textbf{q})$ on the LPT grid in the full box, which is achieved by solving equations \eqref{eq:LPTpotential1} and \eqref{eq:LPTpotential2}. 
\end{enumerate}
 
If initial phases are generated in Fourier space, the Zel'dovich approximation (i.e. the calculation of $\phi^{(1)}$) requires only one inverse fast Fourier transform (FFT) on the LPT grid. For the second-order potential, the source term on the right-hand side of equation \eqref{eq:LPTpotential2} has to be computed from $\phi^{(1)}$; this can either be done in Fourier space (for a cost of six inverse FFTs) or in configuration space via finite differencing (for a cost of nine one-dimensional gradient operations). In both cases, the calculation of $\phi^{(2)}$ from its source then requires one forward and one inverse FFT.

These few FFTs in the full box are the most hardware-demanding requirement of the algorithm (particularly in terms of memory), and the only step which is not distributed and suitable for grid computing. These FFTs may however be performed on a cluster of computers with fast interconnection suitable for Message Passing Interface \citep{FFTW05,Johnson2008FFTW}.

\subsection{Tiling and buffer region}
\label{ssec:Tiling and buffer region}

\begin{enumerate}
\item[B.1.] After having computed the Lagrangian potentials, the master process splits the Lagrangian lattice (of size $N_\mathrm{p}^3$) into $N_\mathrm{tiles}^3$ cubic tiles (we require that $N_\mathrm{p}$ is a multiple of $N_\mathrm{tiles}$). Tiles are constructed to be evolved independently; therefore the main, perfectly parallel region of the algorithm starts here.
\end{enumerate}

To minimise artefacts due to boundary effects (see section \ref{ssec:Evolution of sCOLA boxes}), each tile is surrounded by a ``buffer region'' in Lagrangian space. This buffer region consists of $N_\mathrm{p,buffer}$ particles in each direction, so that each sCOLA box contains a total of $N_\mathrm{p,sCOLA}^3$ particles, where $N_\mathrm{p,sCOLA} \equiv N_\mathrm{p,tile} + 2 N_\mathrm{p,buffer}$ and $N_\mathrm{p,tile} \equiv N_\mathrm{p}/N_\mathrm{tiles}$. Corresponding physical sizes are $L_\mathrm{tile} \equiv L \, N_\mathrm{p,tile}/N_\mathrm{p}$, $L_\mathrm{buffer} \equiv L \, N_\mathrm{p,buffer}/N_\mathrm{p}$, and $L_\mathrm{sCOLA} \equiv L \, N_\mathrm{p,sCOLA}/N_\mathrm{p}$. The fraction of the full Lagrangian lattice assigned to one child sCOLA process is represented by dotted lines in figure \ref{fig:sCOLA_grids}. Particles of the tile are represented in black, and particles of the buffer region are represented in cyan.

The sCOLA box is chosen to encompass the tile and its buffer region. We define the over-simulation factor $r$ as the ratio between the total volume simulated in all sCOLA boxes and the target simulation volume, i.e.
\begin{eqnarray}
r & \equiv & \frac{N_\mathrm{tiles}^3 N_\mathrm{p,sCOLA}^3}{N_\mathrm{p}^3} = \frac{N_\mathrm{tiles}^3 (N_\mathrm{p,tile} + 2N_\mathrm{p,buffer})^3}{N_\mathrm{p}^3}\nonumber\\
& = & \frac{N_\mathrm{tiles}^3 L_\mathrm{sCOLA}^3}{L^3} = \frac{N_\mathrm{tiles}^3 (L_\mathrm{tile} + 2L_\mathrm{buffer})^3}{L^3}.
\end{eqnarray}
Since all sCOLA boxes can be evolved independently, the degree of parallelism of the algorithm is equal to the number of sCOLA boxes, $N_\mathrm{tiles}^3$. We call the ``parallelisation potential factor'' the quantity $p \equiv N_\mathrm{tiles}^3/r$, which balances the degree of parallelism with the amount of over-simulation. It is also
\begin{equation}
p = \frac{N_\mathrm{p}^3}{N_\mathrm{p,sCOLA}^3} = \frac{L^3}{L_\mathrm{sCOLA}^3}.
\end{equation}
For each sCOLA box, the corresponding child process computes the set of final positions $\left\lbrace \textbf{x} \right\rbrace_\mathrm{sCOLA}$ and momenta $\left\lbrace \textbf{p} \right\rbrace_\mathrm{sCOLA}$.

\begin{enumerate}
\item[B.2.] At the end of the evolution, each child process sends the set of final positions $\left\lbrace \textbf{x} \right\rbrace_\mathrm{tile}$ and momenta $\left\lbrace \textbf{p} \right\rbrace_\mathrm{tile}$ of particles of the tile back to the master process. Particles of the buffer region are ignored. The master process then ``untiles'' the simulation by gathering the results from all the tiles.
\end{enumerate}

\subsection{Initial operations in the sCOLA boxes}
\label{ssec:Initial operations in the sCOLA boxes}

A few steps are required in each sCOLA box before starting the evolution per se.

\begin{enumerate}
\item[C.1.] The sCOLA box receives the relevant portion of $\phi^{(1)}(\textbf{q})$ and $\phi^{(2)}(\textbf{q})$ from the master process. This is the only communication required with the master process before sending back the results at the end of the evolution.
\end{enumerate}

The portion of the LPT grid received by each process from the master process corresponds to the full spatial region covered by the sCOLA box, plus an additional padding of $2N_\mathrm{ghost}$ cells in each direction. We denote by $\widetilde{\phi}^{(1)}(\textbf{q})$ and $\widetilde{\phi}^{(2)}(\textbf{q})$ the parts of $\phi^{(1)}(\textbf{q})$ and $\phi^{(2)}(\textbf{q})$ received from the master process (we avoid the superscript ``sCOLA'' since no approximation is involved at this stage). They are defined on a grid of size $(N_\mathrm{sCOLA}+4N_\mathrm{ghost})^3$, where
\begin{equation}
N_\mathrm{tile} \equiv \left\lceil N_\mathrm{p,tile} \frac{N}{N_\mathrm{p}} \right\rceil, \; N_\mathrm{sCOLA} \equiv N_\mathrm{tile} + 2 \left\lceil N_\mathrm{p,buffer} \frac{N}{N_\mathrm{p}} \right\rceil
\end{equation}
($\left\lceil \cdot \right\rceil$ denotes the ceiling function). An illustration is provided in figure \ref{fig:sCOLA_grids}, left panel. There, the portion of the LPT grid corresponding to the sCOLA box, of size $N_\mathrm{sCOLA}$ in each direction, is represented in red and the padding region, of size $2N_\mathrm{ghost}$ in each direction, is represented in pink.

\begin{enumerate}
\item[C.2.] The sCOLA process locally computes the required time-independent LPT vectors $\boldsymbol{\Psi}_1^\mathrm{sCOLA}$ and $\boldsymbol{\Psi}_2^\mathrm{sCOLA}$ via finite differencing in configuration space and interpolation to particles' positions.
\end{enumerate}

The ghost cells included around $\widetilde{\phi}^{(1)}(\textbf{q})$ and $\widetilde{\phi}^{(2)}(\textbf{q})$ in the sCOLA box ensure that the proper boundary conditions are used when applying the gradient operator $\boldsymbol{\nabla}_\textbf{q}^\mathrm{sCOLA}$ in configuration space to get the LPT displacements on the grid. This step ``consumes'' $N_\mathrm{ghost}$ layers of ghost cells in each direction, so that the grid of LPT displacements has a size of $(N_\mathrm{sCOLA} + 2N_\mathrm{ghost})^3$. To use again the proper boundary conditions when going from the LPT grid to particles' positions, another $N_\mathrm{ghost}$ layers of ghost cells is consumed by the interpolation operator $\bar{\mathrm{B}}^\mathrm{sCOLA}$. The use of the exact boundary conditions at each of these two steps ensures that $\boldsymbol{\nabla}_\textbf{q}^\mathrm{sCOLA} = \boldsymbol{\nabla}_\textbf{q}$ and $\bar{\mathrm{B}}^\mathrm{sCOLA} = \bar{\mathrm{B}}$. Therefore, by construction, $\boldsymbol{\Psi}_1^\mathrm{sCOLA} \equiv \boldsymbol{\nabla}_\textbf{q}^\mathrm{sCOLA} \widetilde{\phi}^{(1)}(\textbf{q})$ and $\boldsymbol{\Psi}_2^\mathrm{sCOLA} \equiv \boldsymbol{\nabla}_\textbf{q}^\mathrm{sCOLA} \widetilde{\phi}^{(2)}(\textbf{q})$ in the sCOLA box are always the same as $\boldsymbol{\Psi}_1 \equiv \boldsymbol{\nabla}_\textbf{q} \phi^{(1)}(\textbf{q})$ and $\boldsymbol{\Psi}_2 \equiv \boldsymbol{\nabla}_\textbf{q} \phi^{(2)}(\textbf{q})$ in the full box (as would be computed by the master process). Consequently, we do not keep track of both $\boldsymbol{\Psi}_\mathrm{1,2}^\mathrm{sCOLA}$ and $\boldsymbol{\Psi}_\mathrm{1,2}$, contrary to \citet{Tassev2015}. In addition to being simpler, this scheme has the practical advantage of saving six floating-point numbers per particle in memory (three in the case of the Zel'dovich approximation).

\begin{enumerate}
\item[C.3.] The sCOLA process precomputes the Dirichlet boundary conditions $\Phi_\mathrm{BCs}$ that will be used at each calculation of the gravitational potential during the sCOLA evolution.
\end{enumerate}

For each sCOLA box, we define a particle-mesh grid of size $N_\mathrm{g}^3$ (the ``PM grid'', represented in dark green in the right panel of figure \ref{fig:sCOLA_grids}). The PM grid defines the force resolution; it should be equal to or finer than the LPT grid ($N_\mathrm{g} \geq N_\mathrm{sCOLA}$). Before starting the evolution with sCOLA, each process precomputes the Dirichlet boundary conditions that will be required by the Poisson solver at each value of the scale factor $a_\mathrm{K}$. This calculation takes as input the initial gravitational potential $\widetilde{\phi}^{(1)}(\textbf{q})$ and outputs $\Phi_\mathrm{BCs}(\textbf{x},a_\mathrm{K})$ for each $a_\mathrm{K}$, defined on the PM grid with a padding of $2N_\mathrm{ghost}$ cells around the sCOLA box in each direction (light green and yellow regions in figure \ref{fig:sCOLA_grids}, right panel). The approximation involved in this step is further discussed in section \ref{sssec:Gravitational potential}.

\subsection{Evolution of sCOLA boxes}
\label{ssec:Evolution of sCOLA boxes}

Each sCOLA box is then evolved according to the scheme reviewed in section \ref{ssec:Spatial comoving Lagrangian acceleration (sCOLA)} and appendices \ref{apx:Model equations} and \ref{apx:Standard and modified time-stepping}. Two specific approximations are needed to compute the operators and quantities with a superscript ``sCOLA''; we now discuss the choices that we made.

\subsubsection{Density assignment ($\mathrm{B}^\mathrm{sCOLA}$)}
\label{sssec:Density asssignement}

As mentioned in section \ref{ssec:Spatial comoving Lagrangian acceleration (sCOLA)}, only particles of the sCOLA box should contribute to $\delta^\mathrm{sCOLA}(\textbf{x},a)$. For particles that are fully in the sCOLA box, density assignment can be chosen as the same operation as would be used in a PM or tCOLA code (typically, a CiC scheme). A question is what to do with particles that have (partially) left the sCOLA box during the evolution, while keeping the requirement of no communication between boxes: this constitutes the only difference between the operators $\mathrm{B}$ and $\mathrm{B}^\mathrm{sCOLA}$. Possible choices include artificially periodising the sCOLA box (which is clearly erroneous) or stopping particles at its boundaries (which does not conserve momentum). Both of these choices assign the entire mass carried by the set of sCOLA particles $\mathcal{S}$ to the PM grid, but result in artefacts in the final conditions, if the buffer region is not large enough.

An alternative choice is simply to limit the (Eulerian) PM grid volume where we compute $\delta^\mathrm{sCOLA}(\textbf{x},a)$ to the (Lagrangian) sCOLA box, including central and buffer regions.  In practice, this means ignoring the fractional particle masses that the CiC assignment would have deposited to grid points outside the sCOLA box. We have found in our tests that this choice gives the smallest artefacts of the three choices considered.\footnote{There is a certain symmetry to this choice, since particles that would have moved into the buffer region from the outside are also neglected in the force calculation, due to the lack of communication between different sCOLA boxes.} We note that (partially) erasing some particles' mass is an approximation that is only used in the $\mathrm{B}^\mathrm{sCOLA}$ operator to evaluate the source term in the Poisson equation, and therefore  only affects the force calculation. The number of particles, both within each sCOLA process ($N_\mathrm{p,sCOLA}^3$) and in the full simulation ($N_\mathrm{p}^3$), is left unchanged during the evolution. Therefore, mass is always conserved both within each sCOLA process and within the full volume.

\subsubsection{Gravitational potential ($\Delta_\textbf{x}^\mathrm{sCOLA}$, $\boldsymbol{\nabla}_\textbf{x}^\mathrm{sCOLA}$ and $\bar{\mathrm{B}}^\mathrm{sCOLA}$)}
\label{sssec:Gravitational potential}

\paragraph{Poisson solver ($\Delta_\textbf{x}^\mathrm{sCOLA}$)}

To make sure that differences between $\Phi^\mathrm{sCOLA}(\textbf{x},a)$ and $\Phi(\textbf{x},a)$ are as small as possible, we make use of a Poisson solver with Dirichlet boundary conditions, instead of assuming periodic boundary conditions. Such a Poisson solver uses discrete sine transforms (DSTs) instead of FFTs, and requires the boundary values of $\Phi$ in six planes (west, east, south, north, bottom, top) surrounding the PM grid (see appendix \ref{apx:Poisson solver with Dirichlet boundary conditions}). These planes have a thickness of $N_\mathrm{ghost}$ cells (depending on the value of the FDA used to approximate the Laplacian); they are represented by hatched regions in figure \ref{fig:sCOLA_grids}, right panel. At each scale factor $a_\mathrm{K}$ when the computation of accelerations is needed, the Dirichlet boundary conditions are extracted from the precomputed $\Phi_\mathrm{BCs}(\textbf{x},a_\mathrm{K})$ (step C.3., see section \ref{ssec:Initial operations in the sCOLA boxes}).

Ideally, $\Phi_\mathrm{BCs}(\textbf{x},a_\mathrm{K})$ should be the exact, non-linear gravitational potential in the full volume at $a_\mathrm{K}$, $\Phi(\textbf{x},a_\mathrm{K})$. However, knowing this quantity would require having previously run the monolithic simulation in the full volume, which we seek to avoid. In this paper, we rely instead on the linearly-evolving potential (LEP) approximation \citep{Brainerd1993,Bagla1994}, namely
\begin{equation}
\Phi_\mathrm{BCs}(\textbf{x},a_\mathrm{K}) \approx \Phi_\mathrm{LEP}(\textbf{x},a_\mathrm{K}) \equiv D_1(a_\mathrm{K}) \, \widetilde{\phi}^{(1)}(\textbf{x}) .
\label{eq:LEP}
\end{equation}
The idea behind this approximation is that the gravitational potential is dominated by long-wavelength modes, and therefore it ought to obey linear perturbation theory to a better approximation than the density field.

In equation \eqref{eq:LEP}, we have assumed that the linear growth factor $D_1$ is normalised to unity at the scale factor corresponding to the initial conditions. The precomputation of $\Phi_\mathrm{BCs}$ in step C.3. is therefore an interpolation from the LPT grid to the PM grid and a simple scaling with $D_1(a_\mathrm{K})$.

The output of the Poisson solver is the gravitational potential $\Phi^\mathrm{sCOLA}(\textbf{x},a_\mathrm{K})$ on the PM grid, in the interior of the sCOLA box (dark green grid points in figure \ref{fig:sCOLA_grids}, right panel). Consistently with the treatment above, $\Phi^\mathrm{sCOLA}(\textbf{x},a_\mathrm{K})$ is padded using the values of $\Phi_\mathrm{BCs}(\textbf{x},a_\mathrm{K})$ in $2N_\mathrm{ghost}$ cells around the PM grid, in each direction (light green and yellow regions in figure \ref{fig:sCOLA_grids}, right panel).

Therefore, the only difference between $\Delta_\textbf{x}^\mathrm{sCOLA}$ and $\Delta_\textbf{x}$ resides in using the LEP instead of the true, non-linear gravitational potential at the boundaries of the sCOLA box.

\paragraph{Accelerations ($\boldsymbol{\nabla}_\textbf{x}^\mathrm{sCOLA}$ and $\bar{\mathrm{B}}^\mathrm{sCOLA}$)}

Given the gravitational potential $\Phi^\mathrm{sCOLA}(\textbf{x},a_\mathrm{K})$, accelerations are computed by finite differencing in configuration space and interpolation to particles' positions, similarly to step C.2. (see section \ref{ssec:Initial operations in the sCOLA boxes}). The application of $\boldsymbol{\nabla}_\textbf{x}^\mathrm{sCOLA}$ consumes $N_\mathrm{ghost}$ cells, so that accelerations are obtained on the PM grid with a padding of $N_\mathrm{ghost}$ cells (yellow region in figure \ref{fig:sCOLA_grids}, right panel). Interpolation from the grid to particles' position (the $\bar{\mathrm{B}}^\mathrm{sCOLA}$ operator) further consumes $N_\mathrm{ghost}$ cells.

As for the Laplacian, the only difference between $\boldsymbol{\nabla}_\textbf{x}^\mathrm{sCOLA}$ and $\boldsymbol{\nabla}_\textbf{x}$, and $\bar{\mathrm{B}}^\mathrm{sCOLA}$ and $\bar{\mathrm{B}}$, resides in using the LEP in $\Phi^\mathrm{sCOLA}(\textbf{x},a_\mathrm{K})$ instead of the true, non-linear gravitational potential at the boundaries of the sCOLA box.

\section{Accuracy and speed}
\label{sec:Accuracy and speed}

\begin{table*}
	\centering
	\begin{tabular}{ccccccccccc}
	\hline\hline
    $L$ [$\mathrm{Mpc}/h$] & $N_\mathrm{p}$ & $N$ & $N_\mathrm{tiles}$ & $N_\mathrm{p,tile}$ & $L_\mathrm{tile}$ [$\mathrm{Mpc}/h$] & $N_\mathrm{p,buffer}$ & $L_\mathrm{buffer}$ [$\mathrm{Mpc}/h$] & $N_\mathrm{g}$ & $r$ & $p$ \\
	\hline
    200 & 512 & 256 & 16 & 32 & 12.5 & 32 & 12.5 & 97 & 27 & 151.70 \\
     &  &  & 8 & 64 & 25 & 32 & 12.5 & 129 & 8 & 64 \\
     &  &  & 8 & 64 & 25 & 64 & 25 & 193 & 27 & 18.96 \\
     &  &  & 4 & 128 & 50 & 32 & 12.5 & 193 & 3.38 & 18.96 \\
     &  &  & 4 & 128 & 50 & 64 & 25 & 257 & 8 & 8 \\
     &  &  & 4 & 128 & 50 & 128 & 50 & 385 & 27 & 2.37 \\
     &  &  & 2 & 256 & 100 & 32 & 12.5 & 321 & 1.95 & 4.10 \\
     &  &  & 2 & 256 & 100 & 64 & 25 & 385 & 3.38 & 2.37 \\
	\hline
    1000 & 1024 & 512 & 16 & 64 & 62.5 & 14 & 13.7 & 93 & 2.97 & 1378.91 \\
    & & & 16 & 64 & 62.5 & 26 & 25.4 & 117 & 5.95 & 687.90 \\
    & & & 16 & 64 & 62.5 & 40 & 39.1 & 145 & 11.39 & 359.59 \\
    & & & 16 & 64 & 62.5 & 64 & 62.5 & 193 & 27 & 151.70 \\
    & & & 8 & 128 & 125 & 10 & 9.8 & 149 & 1.55 & 331.22 \\
    & & & 8 & 128 & 125 & 20 & 19.5 & 169 & 2.26 & 226.45 \\
    & & & 8 & 128 & 125 & 30 & 29.3 & 189 & 3.17 & 161.59 \\
    & & & 8 & 128 & 125 & 50 & 48.8 & 229 & 5.65 & 90.59 \\
	\hline\hline
	\end{tabular}
	\caption{Different setups used to test the accuracy and speed of our sCOLA algorithm.}
	\label{tb:simulations}
\end{table*}

We implemented the perfectly parallel sCOLA algorithm described in section \ref{sec:Algorithm for perfectly parallel simulations using sCOLA} in the \textsc{Simbelmynë} code \citep{Leclercq2015ST}, publicly available at \url{https://bitbucket.org/florent-leclercq/simbelmyne/} \citep[see also][appendix B, for technical details on the implementation of the PM and tCOLA models in \textsc{Simbelmynë}]{LeclercqThesis}. This section describes some tests of the accuracy and speed of the new sCOLA algorithm. Since our implementation, relying on evaluating forces with a PM scheme, introduces some additional approximations with respect to tCOLA, we compare our results to that of corresponding monolithic tCOLA simulations. The accuracy of tCOLA with respect to more accurate gravity solvers has been characterised in the earlier literature \citep{Tassev2013,Howlett2015,Leclercq2015ST,Koda2016,Izard2016}. The question of comparing the accuracy of our sCOLA algorithm to full $N$-body simulations would require building in a full $N$-body integrator for the sCOLA boxes (see equations \eqref{eq:sCOLA_NbodyEoM} and \eqref{eq:force_sCOLA}); this subject is left for future research.

Throughout the paper, we adopt the $\Lambda$CDM model with Planck 2015 cosmological parameters: $h=0.6774$, $\Omega_\Lambda = 0.6911$, $\Omega_\mathrm{b} = 0.0486$, $\Omega_\mathrm{m} = 0.3089$, $n_\mathrm{S} = 0.9667$, $\sigma_8 = 0.8159$ \citep[][page 31, table 4, last column]{PlanckCollaboration2015}. The initial power spectrum is computed using the \citet{Eisenstein1998,Eisenstein1999} fitting function.

\interfootnotelinepenalty=10000
We base our first tests on a periodic box of comoving side length $L=200$~$\mathrm{Mpc}/h$ populated with $N_\mathrm{p}^3 = 512^3$ dark matter particles. For all operators, we use FDA at order 2. The LPT grid has $N^3 = 256^3$ voxels. Particles are evolved to redshift $z=19$ using 2LPT. For all runs, we use $10$ time-steps linearly-spaced in the scale factor to evolve particles from $z=19$ ($a_\mathrm{i}=0.05$) to $z=0$ ($a_\mathrm{f}=1$) (see appendix \ref{apx:Standard and modified time-stepping}).\footnote{This means that in the case of our new sCOLA algorithm, we use COLA both ``in space and time'' \citep[see][]{Tassev2015}.} For tCOLA, the PM grid, covering the full box, has $512^3$ voxels. For sCOLA, we use eight different setups, with various parameters $\{N_\mathrm{tiles}, N_\mathrm{p,tile}, L_\mathrm{tile}, N_\mathrm{p,buffer}, L_\mathrm{buffer}, N_\mathrm{g}, r, p \}$ given in the first part of table \ref{tb:simulations}.

To assess more extensively the impact of using sCOLA on large scales, we used a second ensemble of simulations with the following differences: a box with comoving side length of $L=1$~$\mathrm{Gpc}/h$, $N_\mathrm{p}=1024^3$ particles, a LPT grid with $N^3 = 512^3$ voxels, and a PM grid of $1024^3$ voxels for tCOLA. For sCOLA, we use eight different setups given in the second part of table \ref{tb:simulations}.

\subsection{Qualitative assessments}
\label{ssec:Qualitative assessments}

\begin{figure*}[!thp]
\includegraphics[width=\linewidth]{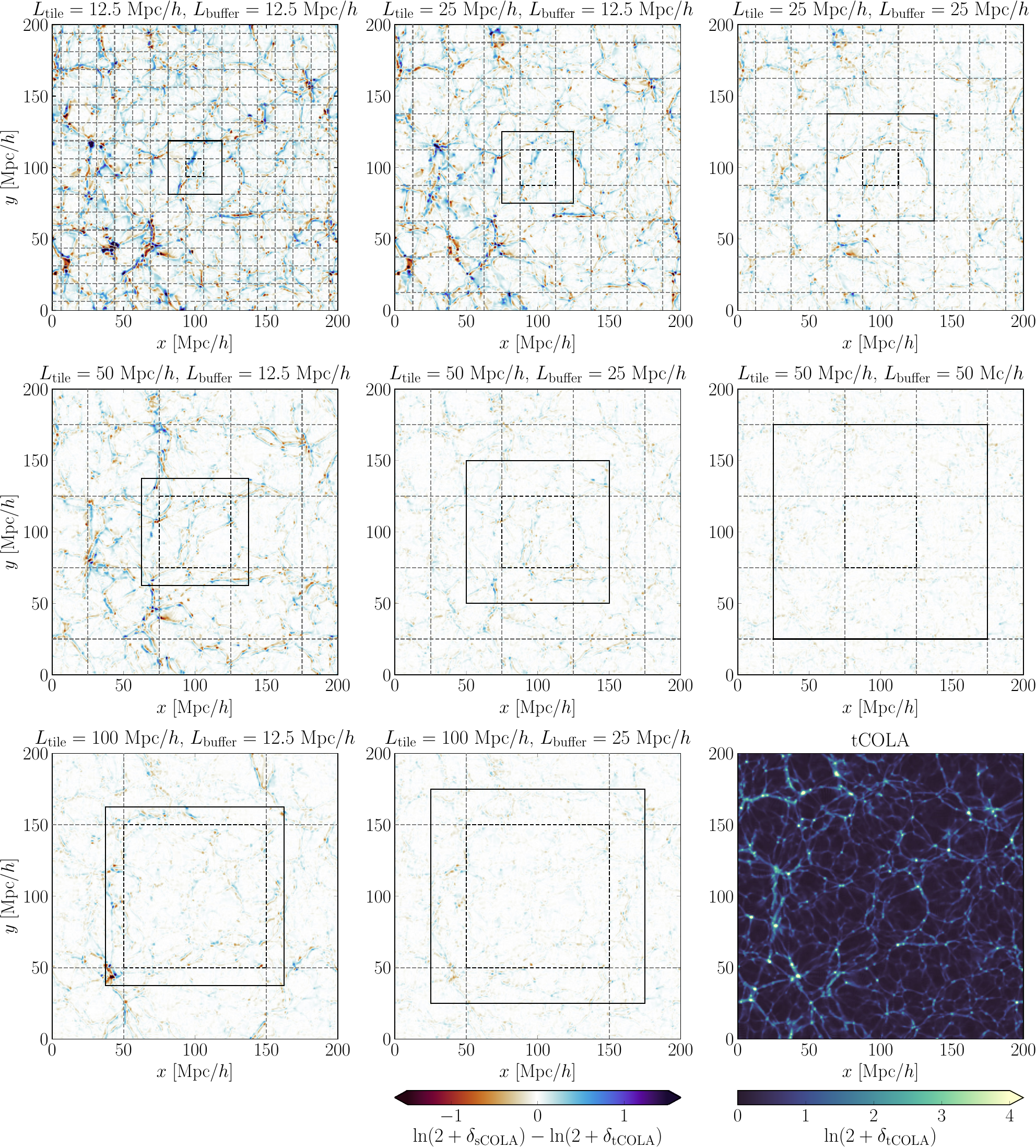}
\caption{Qualitative assessment of the redshift-zero density field from sCOLA for different tilings and buffer sizes, with respect to tCOLA. The bottom right panel shows the reference tCOLA density field in a $200$~$\mathrm{Mpc}/h$ box with periodic boundary conditions (the quantity represented is $\ln(2+\delta_\mathrm{tCOLA})$ where $\delta_\mathrm{tCOLA}$ is the density contrast). Other panels show the difference between sCOLA and tCOLA density fields, $\ln(2+\delta_\mathrm{sCOLA})-\ln(2+\delta_\mathrm{tCOLA})$, for different sizes of tile and buffer region, as indicated above the panels. The tiling is represented by dashed lines, and the central tile's buffer region is represented by solid lines. In the third dimension, the slices represented intersect the central tile at its centre. As can be observed in this figure, artefacts are predominantly located close to the boundaries of tiles; they are reduced with increasing tile size and buffer region size.}
\label{fig:200Mpc_density}
\end{figure*}

The redshift-zero density field is estimated by assigning all particles to the LPT grid using the CiC scheme. Results for the $200$~$\mathrm{Mpc}/h$ box are shown in figure \ref{fig:200Mpc_density}. There, the bottom right panel shows the reference tCOLA density field and other panels show the differences between sCOLA and tCOLA results, for the eight different setups. Some qualitative observations can be made: when artefacts are visible in the sCOLA results, they mainly affect over-dense regions of the cosmic web (filaments and halos), whereas under-dense regions are generally better recovered. Artefacts are of two types: the position of a structure (usually a filament) can be imprecise due to a misestimation of bulk motions (this is visible as a ``dipole'' in figure \ref{fig:200Mpc_density}); or the density (usually of halos) can be over- or under-estimated (this is visible as a ``monopole'' in figure \ref{fig:200Mpc_density}). In all setups, artefacts are predominantly located close to the boundaries of tiles (represented as dashed lines) and are less visible in the centre of tiles. This can be easily understood given that the approximations made all concern the behaviour at the boundaries of sCOLA boxes. At fixed size for the buffer region, the correspondence between sCOLA and tCOLA density fields improves with increasing tile size. A minimum tile size of about $50$~$\mathrm{Mpc}/h$ seems necessary to limit the misestimation of halo densities (``monopoles'' in figure \ref{fig:200Mpc_density}). At low redshift, this scale is in the mildly non-linear regime, where LPT starts to break down; therefore, the LPT frame is inaccurate for particles, and the requirement of no communication between tiles leads to mispredicted clustering. As expected, at fixed tile size, the results are improved by increasing the buffer region around tiles: in each sCOLA box, boundary approximations are pushed farther away from the central region of interest. A good compromise between reducing artefacts and increasing the size of buffer regions seems to be found for a buffer region of $25$~$\mathrm{Mpc}/h$, which corresponds roughly to the maximum distance travelled by a particle from its initial to its final position. In particular, the setup $L_\mathrm{tile} = 50$~$\mathrm{Mpc}/h$, $L_\mathrm{buffer}=25$~$\mathrm{Mpc}/h$ leads to a satisfactory approximation of the tCOLA density with a parallelisation potential factor $p=8$.

\begin{figure*}[!th]
\includegraphics[width=\textwidth]{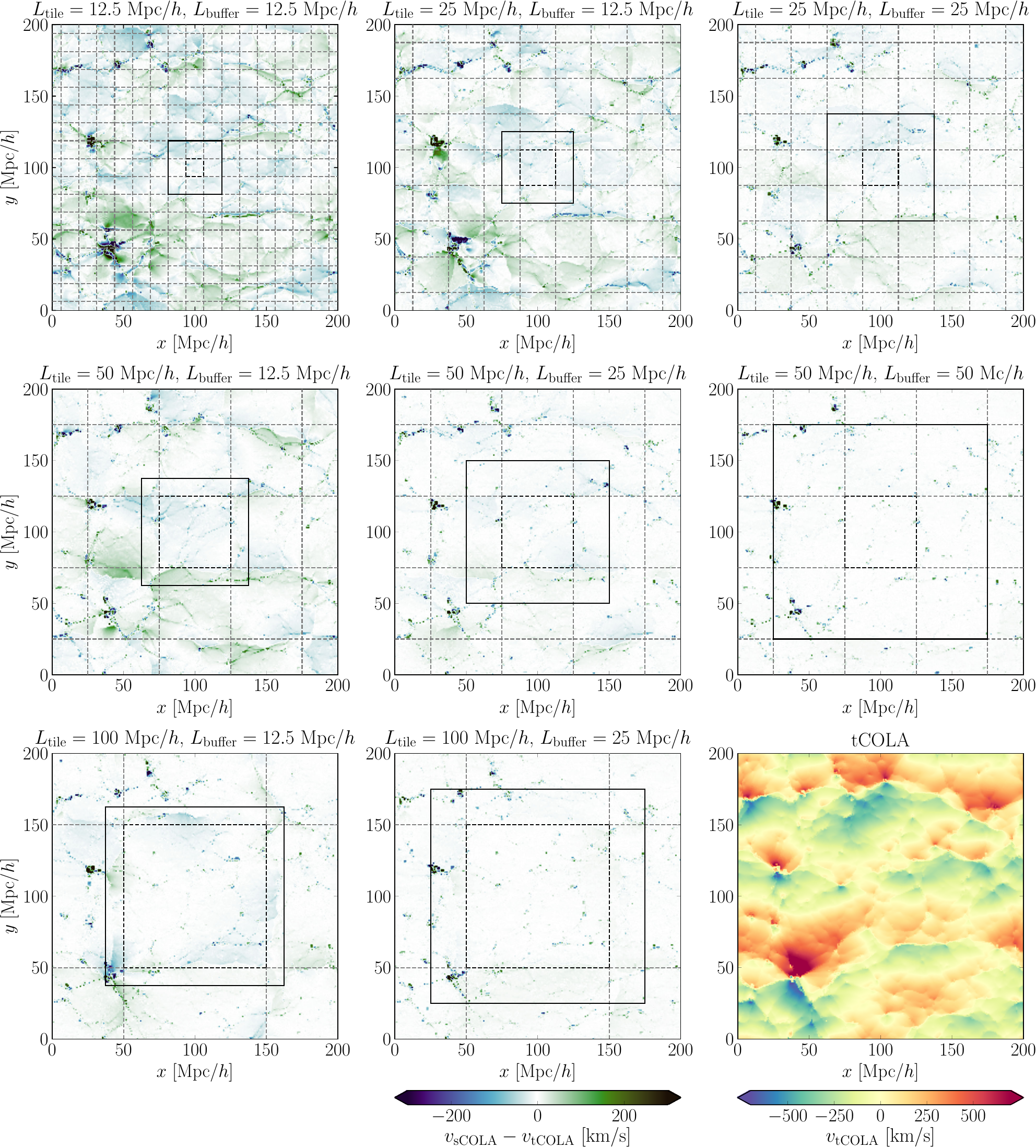} 
\caption{Same as figure \ref{fig:200Mpc_density}, but for one component of the velocity field, in km/s. Bulk flows are correctly captured if tiles and their buffer regions are large enough. Residual differences inside halos can be observed, but they are expected due to the limited number of time-steps, rendering both tCOLA and sCOLA velocities inaccurate in the deeply non-linear regime.}
\label{fig:200Mpc_velocity}
\end{figure*}

{In a similar fashion, the velocity field is estimated on the LPT grid from particle information, using the simplex-in-cell estimator \citep{HahnAnguloAbel2015,Leclercq2017DMSHEET}. Using phase-space information, this estimator accurately captures the velocity field, even in regions sparsely sampled by simulation particles. Results for the $200$~$\mathrm{Mpc}/h$ box are shown in figure \ref{fig:200Mpc_velocity}, where one component of the tCOLA velocity field $v_\mathrm{tCOLA}$ (in km/s) is shown in the bottom right panel. Other panels \parfillskip=0pt\par}

\clearpage 
\onecolumngrid
\clearpage

\twocolumngrid

\interfootnotelinepenalty=100

\noindent show the velocity error in sCOLA, $v_\mathrm{sCOLA}-v_\mathrm{tCOLA}$ in km/s. Differences between tCOLA and sCOLA velocity fields are of two kinds: misestimation of bulk flows (visible as light, spatially extended regions in figure \ref{fig:200Mpc_velocity}), or misestimation of particle velocities inside halos (visible as dark spots in figure \ref{fig:200Mpc_velocity}). We do not interpret the second kind of differences as errors made by our sCOLA algorithm: indeed, motions within virialised regions are not captured accurately by any simulation using only ten time-steps, even by tCOLA in the full box. Therefore, only the first kind of differences, that is, the misestimation of coherent bulk motions is physically interpretable. In this respect, the same behaviour as for density fields can be observed: artefacts are mostly located at the boundaries of tiles, and they are reduced with increasing tile size and buffer region size, with safe minima of $L_\mathrm{tile} \gtrsim 50$~$\mathrm{Mpc}/h$ and $L_\mathrm{buffer} \gtrsim 25$~$\mathrm{Mpc}/h$, respectively.

\subsection{Summary statistics}
\label{ssec:Summary statistics}

In this section, we turn to a more quantitative assessment of our results, by checking the power spectrum of final density fields and their cross-correlation to the tCOLA density field. Even if final density fields are non-Gaussian, two-point statistics (auto- and cross-spectra) are expected to be sensitive to the approximations made in our sCOLA algorithm, which involves both local and non-local operations in configuration space.

\begin{figure}
\includegraphics[width=\linewidth]{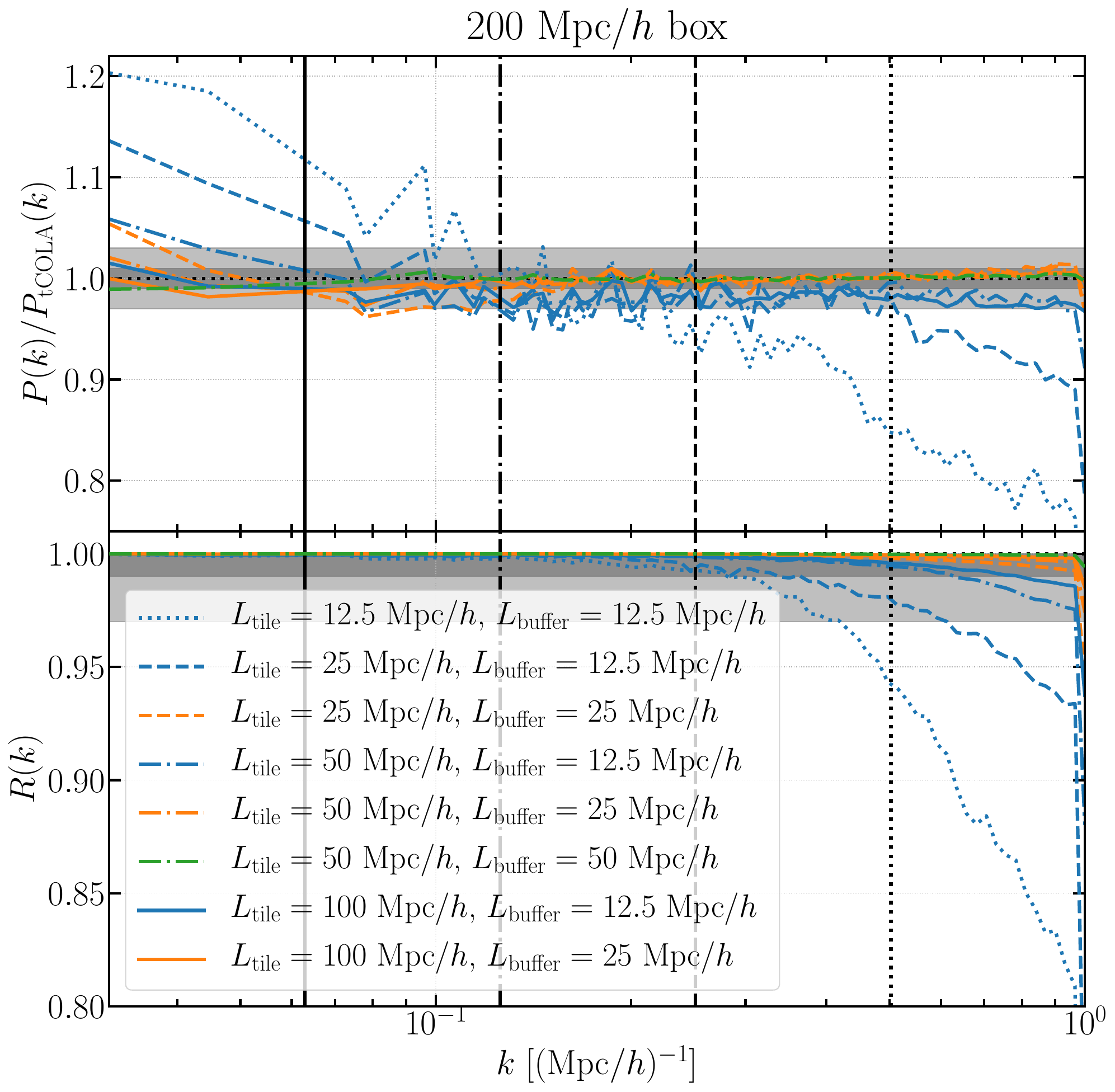} 
\caption{Power spectrum relative to tCOLA (top panel) and cross-correlation with respect to tCOLA (bottom panel) of redshift-zero sCOLA density fields, in a $200~\mathrm{Mpc}/h$ box containing $512^3$ dark matter particles. Different sizes for the tiles (represented by different line styles) and buffer regions (represented by different colours) are used, as indicated in the legend. The vertical lines show the respective fundamental mode of different tiles, the light grey bands correspond to $3\%$ accuracy, and the dark grey bands to $1\%$ accuracy.}
\label{fig:200Mpc_power}
\end{figure}

\begin{figure}
\includegraphics[width=\linewidth]{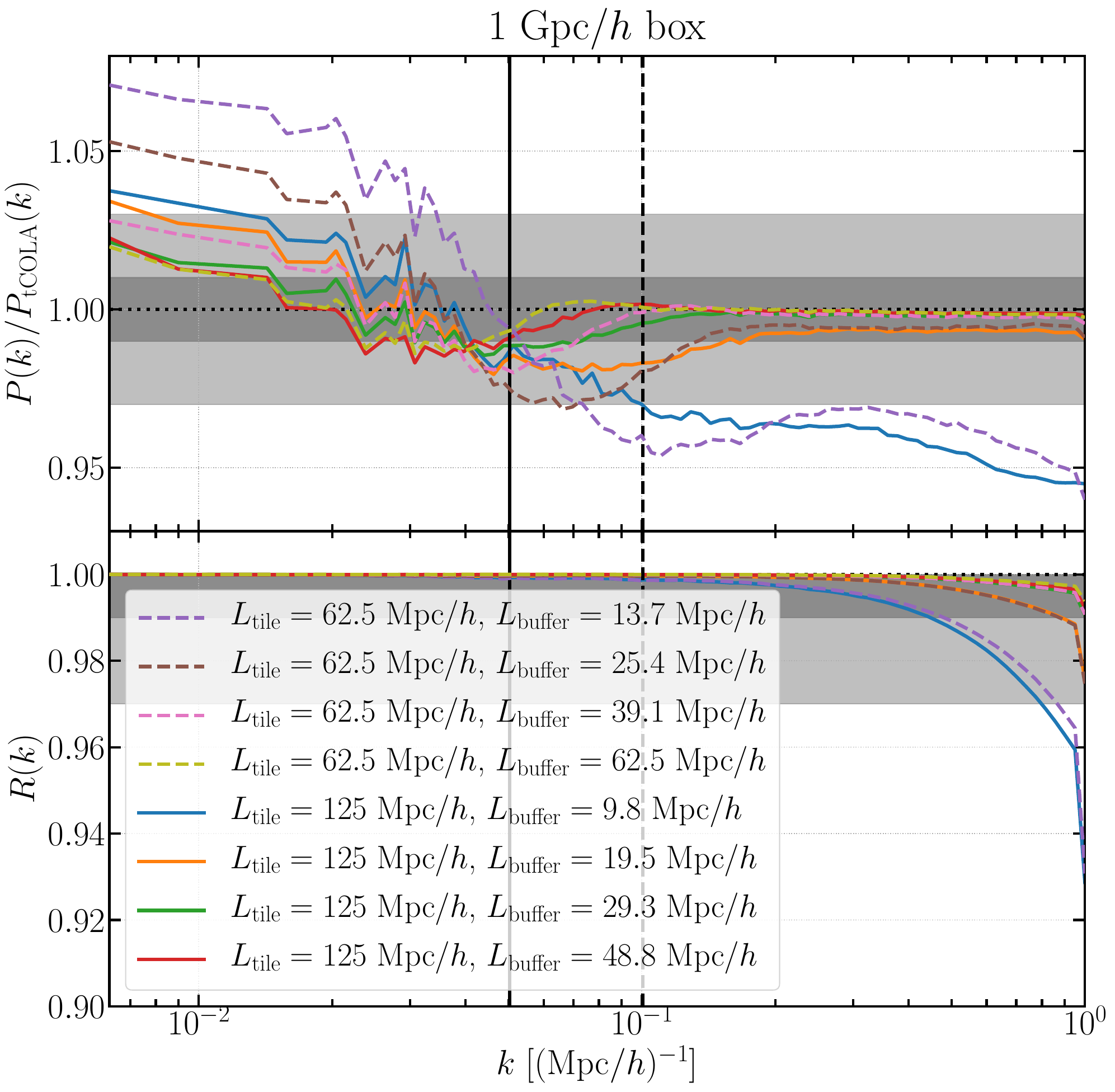} 
\caption{Same as figure \ref{fig:200Mpc_power}, but in a $1~\mathrm{Gpc}/h$ box containing $1024^3$ particles.}
\label{fig:1Gpc_power}
\end{figure}

According to \citet{Huterer2005} or \citet{Audren2013}, in the best cases, observational errors for a Euclid-like survey are typically of order $3\%$ for $k < 10^{-2}~(\mathrm{Mpc}/h)^{-1}$. These results do not account for any of the systematic uncertainties linked to selection effects or contamination of the clustering signal by foregrounds. At smaller scales, theoretical uncertainties take over, reaching $1\%$ and above for $k > 10^{-1}~(\mathrm{Mpc}/h)^{-1}$. In addition, the impact of baryonic physics is still largely uncertain, some models predicting an impact of at least $10\%$ at $k=1~(\mathrm{Mpc}/h)^{-1}$ \citep[e.g.][]{vanDaalen2011,Chisari2018,Schneider2019}. Any data model involving our sCOLA algorithm will be subject to these uncertainties. For this reason, we aim for no better than $3\%$ to $1\%$ accuracy at all scales up to $k =1$~$(\mathrm{Mpc}/h)^{-1}$, for any two-point measurement of clustering.

More precisely, we work with $P(k)$ and $R(k)$, defined for two density contrast fields $\delta$ and $\delta' = \delta_\mathrm{tCOLA}$, with our Fourier transform convention, by
\begin{eqnarray}
\updelta_\mathrm{D}(\textbf{k}-\textbf{k}') P(k) & \equiv & (2\pi)^{-3} L^6 \left\langle \delta^*(\textbf{k}) \delta(\textbf{k}') \right\rangle, \\
\updelta_\mathrm{D}(\textbf{k}-\textbf{k}') R(k) & \equiv & \frac{\left\langle \delta^*(\textbf{k})\delta'(\textbf{k}') \right\rangle}{\sqrt{\left\langle \delta^*(\textbf{k})\delta(\textbf{k}') \right\rangle \left\langle \delta'^*(\textbf{k})\delta'(\textbf{k}') \right\rangle}},
\end{eqnarray}
where $\updelta_\mathrm{D}$ is a Dirac delta distribution. For the estimation of $P(k)$ and $R(k)$, we use $100$ logarithmically-spaced $k$-bins from the fundamental mode of the box $k_\mathrm{min} \equiv 2\pi/L$ to $k =1$~$(\mathrm{Mpc}/h)^{-1}$. 

In figures \ref{fig:200Mpc_power} and \ref{fig:1Gpc_power}, we plot the power spectrum of sCOLA density fields divided by the power spectrum of the reference tCOLA density field, $P_\mathrm{sCOLA}(k)/P_\mathrm{tCOLA}(k)$ (upper panels) and the cross-correlation between sCOLA and tCOLA density fields, $R(k)$ (bottom panels), for our $200~\mathrm{Mpc}/h$ (figure \ref{fig:200Mpc_power}) and $1~\mathrm{Gpc}/h$ box (figure \ref{fig:1Gpc_power}). The grey horizontal bands represent the target accuracies of $3\%$ and $1\%$, and the vertical lines mark the fundamental modes of the tiles, $k_\mathrm{tile} \equiv 2\pi/L_\mathrm{tile}$, for the different values of $L_\mathrm{tile}$ used.

Figure \ref{fig:200Mpc_power} quantitatively confirms the considerations of section \ref{ssec:Qualitative assessments}. Both the amplitudes (as probed by $P(k)/P_\mathrm{tCOLA}(k)$) and the phase accuracy (as probed by $R(k)$) of sCOLA simulations are improved with increasing tile size, for a fixed buffer region (different line styles, same colours). For a fixed tile size, results are also improved by increasing the size of the buffer region (same line styles, different colours). Remarkably, all setups yield perfect phase accuracy at large scales ($R(k)=1$ for $k \leq 0.2~(\mathrm{Mpc}/h)^{-1}$), even when the amplitude of corresponding modes deviates from the tCOLA result. Defects at small scales (lack of power and inaccurate phases) are only observed for the smallest tile sizes and are fixed by increasing the size of buffer region. This effect can be interpreted in Lagrangian coordinates: when the Lagrangian volume forming a halo is divided among different tiles that do not exchange particles, and if the buffer region is too small to contain the rest of the halo, the resulting structure is then split and under-clustered in Eulerian coordinates. In this respect, preferring a sCOLA box size ($L_\mathrm{sCOLA} \equiv L_\mathrm{tile} + 2L_\mathrm{buffer}$) of at least $100~\mathrm{Mpc}/h$ (and therefore $L_\mathrm{tile} \gtrsim 50~\mathrm{Mpc}/h$, $L_\mathrm{buffer} \gtrsim 25~\mathrm{Mpc}/h$, in most situations) seems to be sensible. A more difficult issue is the amplitude of large-scale modes, for $k < k_\mathrm{tile}$. These are sensitive to the tiling if buffer regions around tiles are too small. A safe requirement also seems to be $L_\mathrm{buffer} \gtrsim 25~\mathrm{Mpc}/h$. Putting everything together, in our $200~\mathrm{Mpc}/h$ box, three setups reach $3\%$ accuracy in amplitude and phases at all scales: $\{L_\mathrm{tile} = 50~\mathrm{Mpc}/h, L_\mathrm{buffer} = 25~\mathrm{Mpc}/h\}$ (discussed already in section \ref{ssec:Qualitative assessments}); $\{L_\mathrm{tile} = 100~\mathrm{Mpc}/h, L_\mathrm{buffer} = 25~\mathrm{Mpc}/h\}$; and $\{L_\mathrm{tile} = 50~\mathrm{Mpc}/h, L_\mathrm{buffer} = 50~\mathrm{Mpc}/h\}$. The last-mentioned performs even better, reaching $1\%$ accuracy at all scales, but at the price of over-simulating the volume by a larger factor.

Figure \ref{fig:1Gpc_power} shows the same diagnostics for a $1~\mathrm{Gpc}/h$ box, where the qualitative behaviour is the same as before. It confirms the requirement $L_\mathrm{buffer} \gtrsim 25~\mathrm{Mpc}/h$ to get sufficient accuracy at high $k$. The question of the accuracy reached at the largest scales is then jointly sensitive to $L_\mathrm{tile}$ and $L$. In our tests, the setups $\{L_\mathrm{tile} = 62.5~\mathrm{Mpc}/h, L_\mathrm{buffer} = 39.1~\mathrm{Mpc}/h\}$ and $\{L_\mathrm{tile} = 125~\mathrm{Mpc}/h, L_\mathrm{buffer} = 29.3~\mathrm{Mpc}/h\}$ yield $3\%$ accurate results at all scales, and the setups $\{L_\mathrm{tile} = 62.5~\mathrm{Mpc}/h, L_\mathrm{buffer} = 62.5~\mathrm{Mpc}/h\}$ and $\{L_\mathrm{tile} = 125~\mathrm{Mpc}/h, L_\mathrm{buffer} = 48.8~\mathrm{Mpc}/h\}$ almost reach $1\%$-level precision at all scales. We note that the two different boxes have different mass resolutions, which confirms that requirements for tile and buffer region sizes should be expressed in physical size.

\subsection{Tests of the approximations}
\label{ssec:Tests of the approximations}

\begin{figure}
\includegraphics[width=\linewidth]{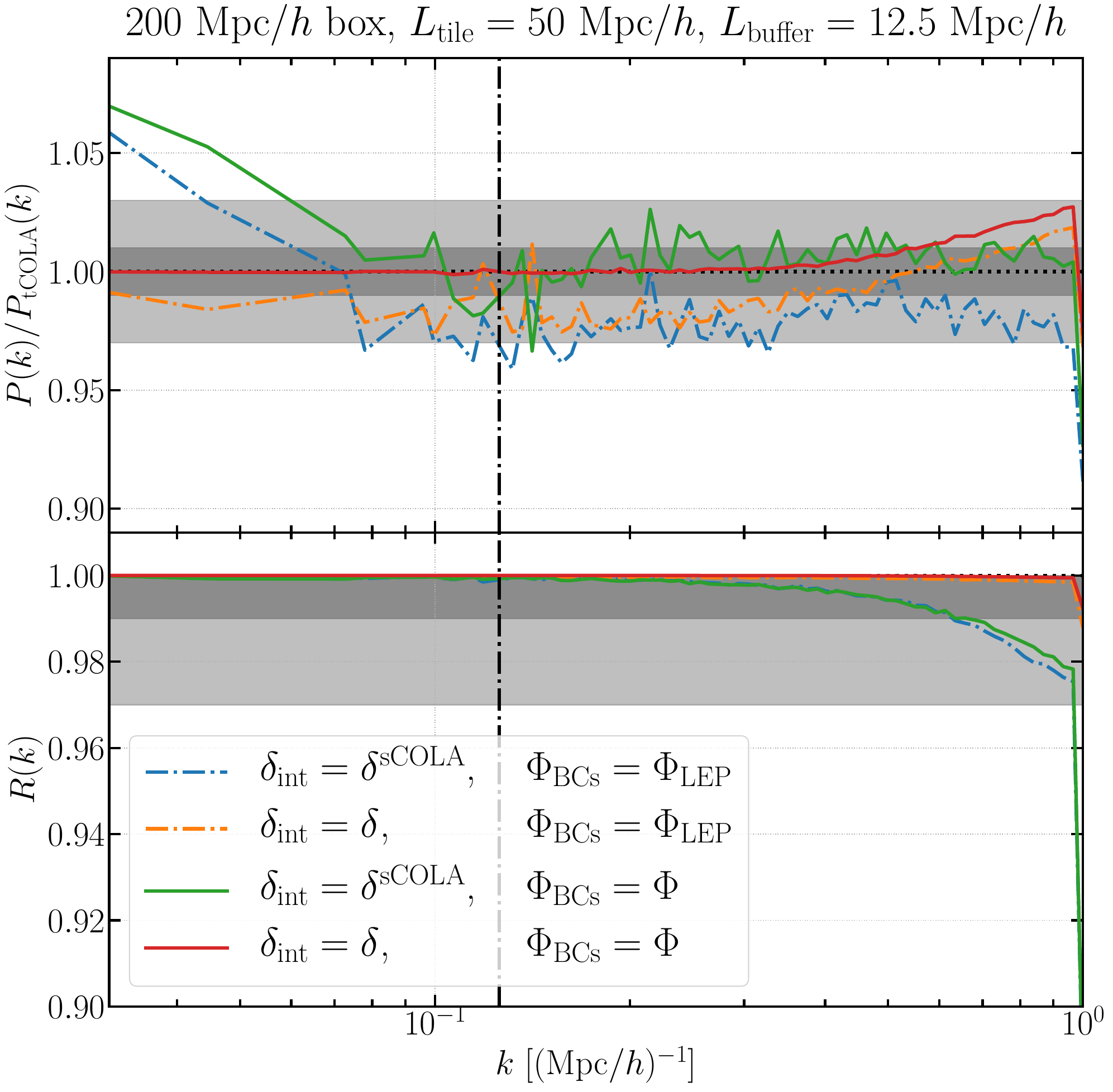} 
\caption{Tests of the approximations made in sCOLA for the density field and the gravitational potential. As in figure \ref{fig:200Mpc_power}, the diagnostic tools are the power spectrum relative to tCOLA (top panel) and the cross-correlation with tCOLA (bottom panel). Our sCOLA algorithm uses the approximate interior density field $\delta^\mathrm{sCOLA}$ and the LEP approximation for the boundary gravitational potential (dash-dotted blue line). In other simulations, as indicated in the legend, we use the true density field $\delta$ and/or the true gravitational potential $\Phi$ at the boundaries. The approximation made for the density field dominates, especially at large scales.}
\label{fig:200Mpc_power_approxs}
\end{figure}

As discussed in section \ref{ssec:Evolution of sCOLA boxes}, two approximations are introduced in our sCOLA algorithm with respect to a monolithic tCOLA approach. These concern density assignment in the interior of sCOLA boxes (approximation \hyperref[sssec:Density asssignement]{D.1.}) and the gravitational potential at the boundaries of sCOLA boxes (approximation \hyperref[sssec:Gravitational potential]{D.2.}). In this section, we test the impact of these approximations on final results, using two-point statistics as diagnostic tools. For this test we use our sCOLA run with $L=200~\mathrm{Mpc}/h$, $N_\mathrm{p}=512^3$, $64$ tiles ($N_\mathrm{tiles}=4$, $N_\mathrm{p,tile} = 128$) and $N_\mathrm{p,buffer}=32$ (i.e. $L_\mathrm{tile} = 50~\mathrm{Mpc}/h$, $L_\mathrm{buffer}=12.5~\mathrm{Mpc}/h$). We choose a small buffer size on purpose, to be sensitive to the approximations made.

Let us denote by $\delta_\mathrm{int}$ the density contrast in the interior of sCOLA boxes and by $\Phi_\mathrm{BCs}$ the gravitational potential at the boundaries of sCOLA boxes. As discussed in section \ref{ssec:Evolution of sCOLA boxes}, our algorithm involves an approximation regarding particles leaving the sCOLA box during the evolution, yielding $\delta^\mathrm{sCOLA}$, and relies on the LEP approximation at the boundaries. It therefore uses
\begin{equation}
\delta_\mathrm{int} = \delta^\mathrm{sCOLA} \quad \mathrm{and} \quad \Phi_\mathrm{BCs} = \Phi_\mathrm{LEP}. \label{eq:test_setup1}
\end{equation}
Everything else being fixed, we ran three investigative sCOLA simulations using respectively,
\begin{alignat}{4}
& \delta_\mathrm{int} = \delta && \quad \mathrm{and} \quad \Phi_\mathrm{BCs} = \Phi_\mathrm{LEP}, \label{eq:test_setup2}\\
& \delta_\mathrm{int} = \delta^\mathrm{sCOLA} && \quad \mathrm{and} \quad \Phi_\mathrm{BCs} = \Phi, \label{eq:test_setup3}\\
& \delta_\mathrm{int} = \delta && \quad \mathrm{and} \quad \Phi_\mathrm{BCs} = \Phi, \label{eq:test_setup4}
\end{alignat}
where $\delta$ is the ``true'' density contrast and $\Phi$ is the ``true'' gravitational potential, extracted at each time-step from the corresponding tCOLA simulation.

Figure \ref{fig:200Mpc_power_approxs} shows the auto- and cross-spectra of resulting sCOLA density fields, with respect to the reference tCOLA result. The use of $\delta_\mathrm{int} = \delta$ yields by construction $R(k)=1$ at all scales, as can be checked from the bottom panel. The setup given by equation \eqref{eq:test_setup4} is rid of the two approximations; it is therefore a consistency check: one should retrieve the tCOLA result if no bias is introduced by the tiling and different Poisson solver. As expected, figure \ref{fig:200Mpc_power_approxs} shows that our implementation recovers the tCOLA result at all scales, with only a small excess of power at $k>0.4~(\mathrm{Mpc}/h)^{-1}$ explained by the slightly higher force resolution of the sCOLA run with respect to tCOLA (the PM grid cell sizes are $0.3886$ and $0.3906~\mathrm{Mpc}/h$, respectively).

The setups given by equations \eqref{eq:test_setup2} and \eqref{eq:test_setup3} allow disentangling the impact of approximations \hyperref[sssec:Density asssignement]{D.1.} and \hyperref[sssec:Gravitational potential]{D.2}. In the standard run (equation \eqref{eq:test_setup1}), averaging over tiles and timesteps, $\sim 0.43\%$ of the $512^3$ particles, all of which belonging to the buffer region, do not deposit all of their mass in the calculation of $\delta^\mathrm{sCOLA}$, but $\sim 76.5\%$ on average. This number only slightly increases with time (from $\sim 0.35\%$ at $a=0.05$ to $\sim 0.47\%$ at $a=1$); in other simulations, we have found that it has a stronger dependence on the mass resolution and on the surface of sCOLA boxes. Regarding the accuracy of the LEP approximation, the ratio of the power spectra of $\Phi-\Phi_\mathrm{LEP}$ and of $\Phi$ goes to zero at early times and large scales, and stays below $12\%$ for all scales with wavenumber $k\leq 2\pi/L_\mathrm{sCOLA}$ at $a=1$. As can be observed in figure \ref{fig:200Mpc_power_approxs}, although using the non-linear gravitational potential instead of the LEP improves both $P(k)$ and $R(k)$ for the final density field at all scales with wavenumber $k > 7 \times 10^{-2}~(\mathrm{Mpc}/h)^{-1}$, it does not remove the $\gtrsim 5\%$ bias in amplitude at the largest scales. On the contrary, using the true density contrast solves this problem and yields a $3\%$ accurate result at all scales, which is remarkable given the small buffer size used in this case (the over-simulation factor is only $r=3.38$).

We conclude from these tests that the approximation made regarding the density field (\hyperref[sssec:Density asssignement]{D.1.}) has more impact than the one regarding the gravitational potential (\hyperref[sssec:Gravitational potential]{D.2.}), especially on the largest modes. This result is consistent with the standard paradigm for structure formation, where the density contrast undergoes severe non-linearity at small scales and late times, while the gravitational potential evolves very little. It also suggests that future improvements of our algorithm should focus on finding a better approximation for $\delta^\mathrm{sCOLA}$, rather than $\Phi_\mathrm{BCs}$.

\subsection{Computational cost}
\label{ssec:Computational cost}

\begin{figure}
\includegraphics[width=\columnwidth]{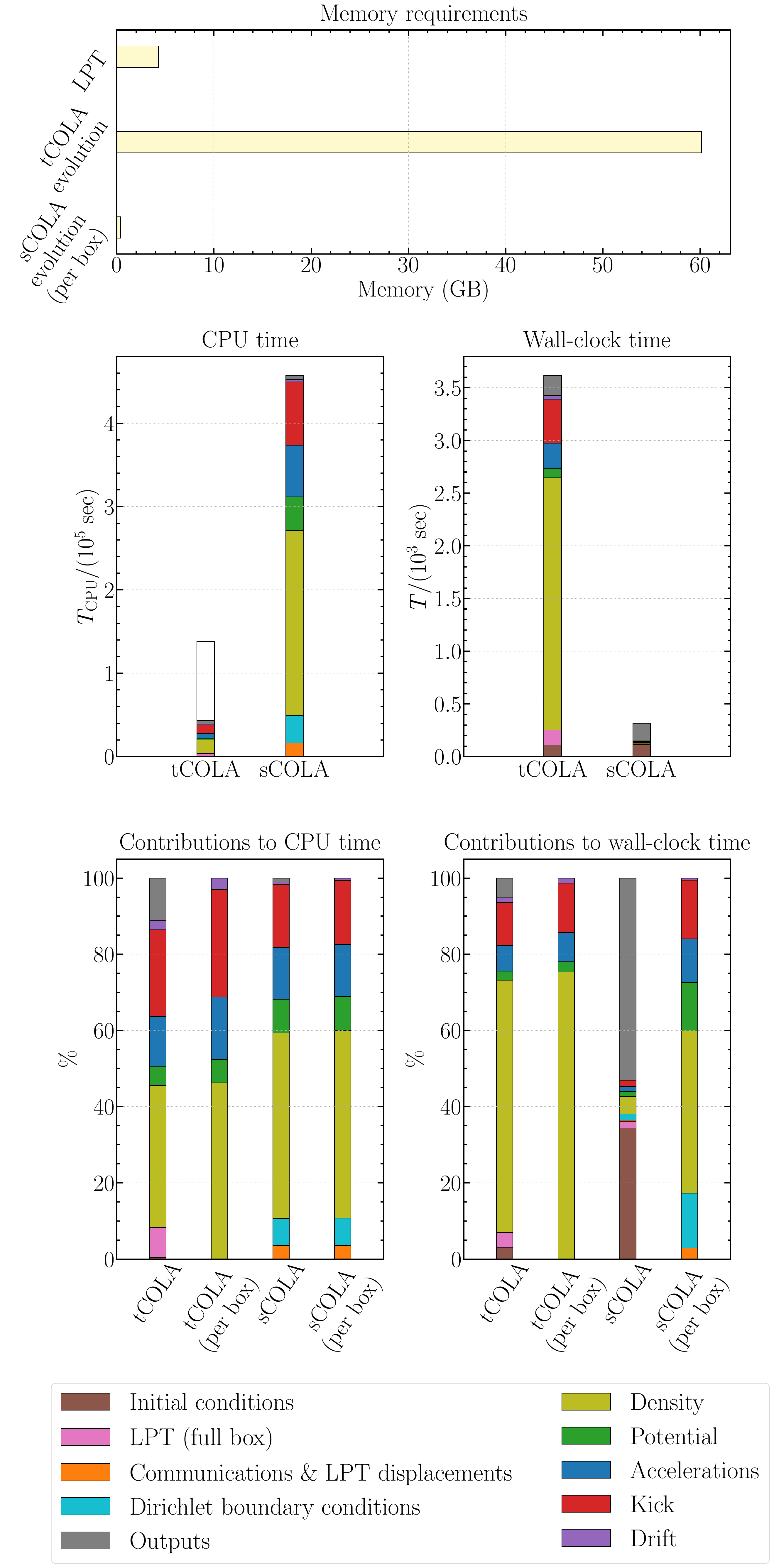} 
\caption{Plots of memory requirements (first row) and of timings for two corresponding tCOLA and sCOLA simulations. Although the CPU time required is higher for sCOLA, the memory consumption and wall-clock time are significantly reduced with respect to tCOLA, due to the perfectly parallel nature of most computations (second row). In the middle left panel, the height of the white bar shows the hypothetical cost of running tCOLA for the same volume as simulated with sCOLA, when taking buffer regions into account. The relative contributions of different operations, as detailed in the legend, is shown in the third row. The main difference in computational cost in sCOLA with respect to tCOLA comes from the use of DSTs instead of FFTs, which makes the evaluation of the potential significantly more expensive.}
\label{fig:timings}
\end{figure}

One of the main motivations for our perfectly parallel algorithm based on sCOLA is to be able to run very large volume simulations at reasonably high resolution. A detailed analysis of the speed and computational cost of our algorithm, as implemented in \textsc{Simbelmynë}, is therefore beyond the intent of this paper. However, in this section we discuss some performance considerations based on a sCOLA run with $L=1~\mathrm{Gpc}/h$, $N_\mathrm{p}=1024^3$, $512$ tiles ($N_\mathrm{tiles}=8$, $N_\mathrm{p,tile} = 128$), $N_\mathrm{p,buffer} = 30$ (i.e. $L_\mathrm{tile} = 125~\mathrm{Mpc}/h$, $L_\mathrm{buffer} = 29.3~\mathrm{Mpc}/h$), $N_\mathrm{g}=199$; and the corresponding monolithic tCOLA simulation. In this case, the over-simulation factor is $r \approx 3.17$ and the parallelisation potential factor is $p \approx 161.59$. To compare the theoretical parallelisation potential factor and the realised parallelisation efficiency, we use one process for tCOLA and $512$ processes for sCOLA. Each process is run on a node with $32$ cores using OpenMP parallelisation.

One of the main advantages of our sCOLA algorithm lies in its reduced memory consumption. In figure \ref{fig:timings} (first row), we show the memory requirements for the calculation of LPT potentials in the full box (common for tCOLA and sCOLA), for the evolution of the full box with tCOLA, and for the evolution of each sCOLA box, all in single-precision floating-point format. LPT requires eight grids of size $N^3$ (one for the initial conditions, one for the Zel'dovich potential, and six for the second-order term), occupying $\sim 4.3$~GB. Evolution with tCOLA requires one integer and $12$ floating-point numbers per particle (their identifier, their position $\textbf{x}$, their momentum $\textbf{p}$, and the vectors $\boldsymbol{\Psi}_1$ and $\boldsymbol{\Psi}_2$), plus a PM grid of $1024^3$ voxels, for a total of $\sim 60.1$~GB. Within each box, sCOLA requires the same memory per particle (but with $N_\mathrm{p,sCOLA}^3 \ll N_\mathrm{p}^3$), a PM grid of size $N_\mathrm{g}^3$, and some overhead for Dirichlet boundary conditions. The total is around $400$~MB per sCOLA box with the setup considered here.

In the second row of figure \ref{fig:timings}, we show the overall cost of tCOLA versus sCOLA, both in terms of CPU time (middle left panel) and wall-clock time (middle right panel). The key feature of our algorithm is that, although the overall CPU time needed is unavoidably higher than with tCOLA, the wall-clock time spent can be drastically reduced. This owes to the degree of parallelism of our algorithm, which is equal to the number of sCOLA boxes. In particular, if as many processes as sCOLA boxes can be allocated ($512$ in this case), the overall wall-clock time is determined by the initial full box operations (common with tCOLA, see section \ref{ssec:Initial conditions and Lagrangian potentials}), plus the cost of evolving only \emph{one} sCOLA box (an average of $30.9$ wall-clock seconds on 32 cores in this test). This is what is shown in the middle right panel of figure \ref{fig:timings}. The wall-clock time reduction factor is $\approx 93$ for the evolution only ($\approx 11$ when accounting for initialisation and writing outputs). Compared to the parallelisation potential factor $p \approx 162$, this number means that sCOLA-specific operations and the larger fractional parallelisation overhead in sCOLA boxes do not significantly hamper the perfectly parallel nature of the code.

The increased CPU time needed with sCOLA (see figure \ref{fig:timings}, middle left panel) is partly due to the necessity of over-simulating the volume of interest by a factor $r>1$ for accuracy. For comparison with the sCOLA CPU time, the height of the white bar shows the tCOLA CPU time multiplied by $r$. The rest of the difference in CPU time principally comes form the fact that simulations with our variant of sCOLA are intrinsically more expensive than with tCOLA for a periodic volume of the same size. This point is further discussed below.

In the third row of figure \ref{fig:timings}, we show the various relative contributions to CPU time and wall-clock time, both for full tCOLA/sCOLA runs and per tCOLA/sCOLA box. The generations of the initial conditions (brown, step \hyperref[ssec:Initial conditions and Lagrangian potentials]{A.1.}) and writing of outputs to disk (grey) are common to tCOLA and sCOLA and have an overall fixed cost. LPT calculations in the full box (pink) consist of computing the Lagrangian potentials and the particle-based LPT displacements in tCOLA, but are limited to computing the Lagrangian potentials in the full box in the case of sCOLA (step \hyperref[ssec:Initial conditions and Lagrangian potentials]{A.2.}). These full-box operations are only showed in the bars labelled ``tCOLA'' and ``sCOLA''. Within each box, the different operations are evaluating the density field (yellow), solving the Poisson equation to get the gravitational potential (green), differentiating the gravitational potential to get the accelerations (blue), ``kicking'' particles (red), and ``drifting'' particles (purple). sCOLA further requires some specific operations within each box: communicating with the master process (steps \hyperref[ssec:Tiling and buffer region]{B.1.}, \hyperref[ssec:Tiling and buffer region]{B.2.}, and \hyperref[ssec:Initial operations in the sCOLA boxes]{C.1.}), calculating the particle-based LPT displacements (step \hyperref[ssec:Initial operations in the sCOLA boxes]{C.2.}), grouped in figure \ref{fig:timings} and shown in orange; and pre-computing the Dirichlet boundary conditions with the LEP approximation (step \hyperref[ssec:Initial operations in the sCOLA boxes]{C.3.}, cyan). sCOLA-specific operations do not contribute more than $10\%$ of the CPU and wall-clock times per box.

A notable difference between evolving a given box with sCOLA or with tCOLA resides in the higher cost of evaluating the potential (green): in this case, $9\%$ of CPU time and $13\%$ of wall-clock time with sCOLA versus $6\%$ of CPU time and $3\%$ of wall-clock time with tCOLA. This effect is due to the use of DSTs, required by the Poisson solver with Dirichlet boundary conditions (see section \ref{ssec:Evolution of sCOLA boxes} and appendix \ref{apx:Poisson solver with Dirichlet boundary conditions}), instead of FFTs. Indeed, depending on the size of the PM grid, the evaluation of DSTs can be the computational bottleneck of our algorithm (up to $60\%$ of overall CPU time is some of our runs), as opposed to the evaluation of the density field (e.g. via CiC) in traditional tCOLA or PM codes ($37\%$ of overall CPU time). For this reason, within each setup, we recommend performing experiments to find a PM grid size giving a good compromise between force accuracy and computational efficiency. In particular, it is strongly preferable that $N_\mathrm{g}+1$ not contain large prime factors (this number appears in the basis functions of sine transforms, see appendix \ref{sapx:Zero-boundary condition Poisson solver}). Throughout this paper, we ensured that $N_\mathrm{g}+1$ is always even, while keeping roughly the same force resolution as the corresponding tCOLA simulation. We note that our choice of $N_\mathrm{g}+1=200$ in the present test, combined with the use of a power of two for the PM grid in the monolithic tCOLA run, favours tCOLA in the comparison of CPU times. The sCOLA CPU time shown in the middle left panel of figure \ref{fig:timings} could be further optimised by making $N_\mathrm{g}+1$ a power of two in sCOLA boxes.

\section{Discussion and conclusion}
\label{sec:Conclusion}

\subsection{Discussion}
\label{ssec:Discussion}

The principal computational challenge of the gravitational $N$-body problem is the long-range nature of the gravitational force. Our sCOLA approach enables perfectly parallel computations and therefore opens up profoundly new possibilities for how to compute large-scale cosmological simulations. We discuss these, some consequences and possible future directions in the following.

\paragraph*{Gravity and physics models}

It is important to note that the sCOLA algorithm introduced in this work is general, and not limited to the gravity model used here: while we focused on a tCOLA particle-mesh implementation to evolve the sCOLA tiles, this choice was designed to facilitate the assessment of tiling artefacts against monolithic tCOLA runs. Nonetheless, any $N$-body method, such as particle-particle--particle-mesh, tree methods or AMR, could be used to evolve each tile. In particular, since the sCOLA approach separates quasi-linear and non-linear scales, there is no need to cut off the computation on small scales. In concert with the approaches discussed below, this fact can be exploited to perform very high-resolution, fully non-linear simulations in cosmological volumes. In this case, the spatial decoupling due to sCOLA would render computations possible that would otherwise be prohibitive.

Similar comments apply to including non-gravitational physics: since hydrodynamical or other non-gravitational forces are typically much more local than gravitational interactions, there are no algorithmic barriers to including them in each sCOLA tile.\footnote{A potential exception is long-range radiative transport of energetic (X-ray or gamma ray) photons, requiring a non-trivial extension of the approach.}

\paragraph*{Construction of light-cones and mock catalogues}

The decoupling of computational volumes achieved by our approach means that each sCOLA box can be run completely independently. Therefore, it is not necessary to define a common final redshift for all tiles. This means that to compute a cosmological light-cone, only a single tile (the one containing the observer) needs to be run to redshift zero. Since the volume on the light-cone increases rapidly with redshift, the vast majority of tiles would only have to be run until they intersect the light-cone at high redshift. In monolithic $N$-body simulations, most of the computational time is spent at low redshift, since the local time-step of simulations decreases with the local dynamical time. Our approach would therefore greatly accelerate the time needed to complete light-cone simulations, by scheduling tiles in order of the redshift to which they should run (and therefore in reverse order of expected computational time), aiding load-balancing.

The construction of light-cones for surveys with large aspect ratios, such as pencil-beam surveys, can further benefit from sCOLA. Indeed, tiles that do not intersect the three-dimensional survey window do not need to be run at all for the construction of mock catalogues. In such a case, the algorithm will still capture the effects of large-scale transverse modes, even if the simulated volume is not substantially increased with respect to the survey volume.

\paragraph*{Low memory requirements}

sCOLA divides the computational volume into much smaller tiles and vastly reduces the memory footprint of each independent sCOLA tile computation, as shown in section \ref{ssec:Computational cost}. As an example, simulating a ${(16~\mathrm{Gpc}/h)}^3$ volume containing $8192^3$ particles to achieve a mass resolution of $10^{12.5}~M_\odot$ requires $\sim 19.8$~TB of RAM with a PM code and $\sim 33.0$~TB of RAM with tCOLA. The setup $\{L_\mathrm{tile} = 62.5~\mathrm{Mpc}/h, L_\mathrm{buffer} = 62.5 ~\mathrm{Mpc}/h\}$ would break down the problem into $256^3$ tiles, each with $(3\times 32)^3$ particles and a memory footprint of $\sim 53~\mathrm{MB}$. This has important consequences, which we explore in the following.

The very modest memory requirement of our algorithm opens up multiple possibilities to accelerate the computation: even on traditional systems, the entire computation of each sCOLA tile would fit entirely into the L3 cache of a multi-core processor. This would cut out the slowest parts of the memory hierarchy, leading to a large potential performance boost and reducing code complexity. Even more promising, many such tiles could be evolved entirely independently on GPU accelerators, or even de\-dicated FPGAs, taking advantage of hybrid architectures of modern computational platforms while reducing the need to develop sophisticated code to manage task parallelism. At this scale, each tile computation would even fit comfortably on ubiquitous small computational platforms such as mobile phones.

\paragraph*{Grid computing}

The perfect scalability achieved by our approach means that large $N$-body simulations can even be run on very inexpensive, strongly asynchronous networks designed for large throughput computing. An extreme example would be participatory computing platforms such as Cosmology@Home,\footnote{\url{https://www.cosmologyathome.org/}} where tens of thousands of users donate computational resources. The use of such platforms would be particularly suited to light-cone computations, as described above. Even if running the low-redshift part necessitates dedicated hardware, other workers could efficiently work independently to compute most of the volume, which lives at high-redshift. Only two communication steps are required for each tile: the LPT potentials are received at the beginning, and at the end of the computation each tile returns its final state at the redshift where it intersects the light-cone.

\paragraph*{Node Failures}

Robustness to node failure is an important consideration on all very large computational platforms. Even with extremely low failure probability for each node, since the number of nodes is high, the probability that some node fails during the course of a computation becomes high. After its initialisation steps (see section \ref{ssec:Initial conditions and Lagrangian potentials}), our approach is entirely robust to such failure, since any individual tile can be recomputed after the fact on a modest system, for very little cost.

\subsection{Conclusion}
\label{ssec:Conclusion}

In this paper, we introduced a perfectly parallel and easily applicable algorithm for cosmological simulations using sCOLA. Our approach is based on a tiling of the full simulation box, where each tile is run independently. By the use of buffer regions and appropriate Dirichlet boundary conditions, we improved the accuracy of the algorithm with respect to \citet{Tassev2015}. In particular, we showed that suitable setups can reach $3\%$ to $1\%$ accuracy at all the scales simulated, as required for data analysis of the next generation of large-scale structure surveys. In case studies, we tested the relative impact of the two approximations involved in our approach, for density assignment and the boundary gravitational potential. We considered the computational cost of our algorithm and demonstrated that even if the CPU time needed is unavoidably higher, the wall-clock time and memory footprint can be drastically reduced.

This study opens up a wide range of possible extensions, discussed in section \ref{ssec:Discussion}. Benefiting from its perfect scalability, the approach could also allow for novel analyses of cosmological data from fully non-linear models previously too expensive to be tractable. It could straightforwardly be used for the construction of mock catalogues, but also within recently introduced likelihood-free inference techniques such as \textsc{delfi} \citep{Alsing2018}, \textsc{bolfi} \citep{Leclercq2018BOLFI} and \textsc{selfi} \citep{Leclercq2019SELFI}, which have a need for cheap simulator-based data models. We therefore anticipate that sCOLA will become an important tool in computational cosmology for the coming era.

Our perfectly parallel sCOLA algorithm has been implemented in the publicly available \textsc{Simbelmynë} code,\footnote{\url{https://bitbucket.org/florent-leclercq/simbelmyne/}} where it is included in version 0.4.0 and later.

\onecolumngrid

\appendix

\section{Model equations}
\label{apx:Model equations}

\subsection{Model equations in the standard PM code}

Denoting by $a$ the scale factor of the Universe and $\tau$ the conformal time, a PM code solves the equations of motion for the position $\textbf{x}$ and momentum $\textbf{p}$ of dark matter particles in comoving coordinates (the mass of particles $m$ is absorbed in the definition of the momentum $\textbf{p}$):
\begin{eqnarray}
\textbf{p} & = & a \deriv{\textbf{x}}{\tau}, \\
\deriv{\textbf{p}}{\tau} & = & - a \boldsymbol{\nabla}_\textbf{x} \Phi(\textbf{x},\tau),
\end{eqnarray}
coupled to the Poisson equation for the gravitational potential, sourced by density fluctuations (equation \eqref{eq:Poisson_full_box}),
\begin{equation}
\Delta_\textbf{x} \Phi(\textbf{x},\tau) = 4\pi \G a^2 \bar{\rho}(\tau) \delta(\textbf{x},\tau),
\label{eq:Poisson-conformal}
\end{equation}
where $\mathrm{G}$ is the gravitational constant and $\bar{\rho}(\tau)$ is the mean matter density at conformal time $\tau$. The density contrast is defined from the local matter density $\rho(\textbf{x},\tau)$ by
\begin{equation}
\delta(\textbf{x},\tau) \equiv \frac{\rho(\textbf{x},\tau)}{\bar{\rho}(\tau)} -1.
\end{equation}
For simplicity, from now on we note $\boldsymbol{\nabla}_\textbf{x} = \boldsymbol{\nabla}$, $\Delta_\textbf{x} = \Delta$ and $\delta(\textbf{x},\tau) = \delta$. 

It is convenient to choose the scale factor as time variable. Using $\partial_\tau = a' \, \partial_a$ and the background evolution $\bar{\rho}(\tau) = \rho^{(0)} a^{-3}$ (a prime denotes a differentiation with respect to $\tau$ and the superscript $(0)$ denotes quantities at the present time), the equations to solve are rewritten:
\begin{eqnarray}
\deriv{\textbf{x}}{a} & = & \frac{\textbf{p}}{a'a} , \\
\deriv{\textbf{p}}{a} & = & -\frac{a}{a'} \boldsymbol{\nabla} \Phi, \label{eq:kick-intermediate}\\
\Delta \Phi & = & 4\pi \G \rho^{(0)} a^{-1} \delta \equiv \frac{3}{2} \Omega_\mathrm{m}^{(0)} \mathcal{H}^{(0)2} a^{-1} \delta . \label{eq:Poisson-intermediate}
\end{eqnarray}

We will use the equivalent formulation
\begin{eqnarray}
\deriv{\textbf{x}}{a} & = & \mathpzc{D}(a) \textbf{p} \quad \mathrm{with} \quad \mathpzc{D}(a) \equiv \frac{1}{a^2 \mathcal{H}(a)}, \label{eq:drift}\\
\deriv{\textbf{p}}{a} & = & \mathpzc{K}(a) \boldsymbol{\nabla} \left( \Delta^{-1} \delta \right) \quad \mathrm{with} \quad \mathpzc{K}(a) \equiv -\frac{3}{2} \frac{\Omega_\mathrm{m}^{(0)} \mathcal{H}^{(0)2}}{a \mathcal{H}(a)}, \label{eq:kick}
\end{eqnarray}
where we have combined equations \eqref{eq:kick-intermediate} and \eqref{eq:Poisson-intermediate}, introduced the conformal Hubble factor $\mathcal{H}(a) \equiv a'/a$, and defined the ``drift prefactor'' $\mathpzc{D}(a)$ and the ``kick prefactor'' $\mathpzc{K}(a)$.

\subsection{Model equations with COLA}

We now introduce the COLA scheme, following \citet{Tassev2013} and \citet{Tassev2015}. For each particle, we work in the frame comoving with its LPT observer, whose position is given by (see section \ref{ssec:Lagrangian perturbation theory (LPT)})
\begin{equation}
\textbf{x}_\mathrm{LPT}(a) = \textbf{q} - D_1(a) \boldsymbol{\Psi}_1 + D_2(a) \boldsymbol{\Psi}_2 ,
\label{eq:LPT-mapping}
\end{equation}
where we have introduced the time-independent vectors $\boldsymbol{\Psi}_1 \equiv \boldsymbol{\nabla}_\textbf{q} \phi^{(1)}$ and $\boldsymbol{\Psi}_2 \equiv \boldsymbol{\nabla}_\textbf{q} \phi^{(2)}$. Noting $\textbf{x}(a) = \textbf{x}_\mathrm{LPT}(a) + \textbf{x}_\mathrm{res}(a)$ the final position of the same particle, we have
\begin{equation}
\deriv{\textbf{x}}{a} = \deriv{\textbf{x}_\mathrm{LPT}}{a} + \deriv{\textbf{x}_\mathrm{res}}{a} ; \quad \mathrm{with} \quad \deriv{\textbf{x}_\mathrm{LPT}}{a} = - \deriv{D_1}{a} \boldsymbol{\Psi}_1 + \deriv{D_2}{a} \boldsymbol{\Psi}_2 \equiv \mathpzc{D}(a) \textbf{p}_\mathrm{LPT} .
\end{equation}
We also define $\textbf{p}_\mathrm{res}$ such that $\drm \textbf{x}_\mathrm{res}/\drm a \equiv \mathpzc{D}(a) \textbf{p}_\mathrm{res}$. Then $\textbf{p} = \textbf{p}_\mathrm{LPT} + \textbf{p}_\mathrm{res}$ (see equation \eqref{eq:drift}). Furthermore,
\begin{equation}
\deriv{\textbf{p}_\mathrm{LPT}}{a} = \frac{\drm}{\drm a} \left( \frac{1}{\mathpzc{D}(a)} \deriv{\textbf{x}_\mathrm{LPT}}{a} \right) \equiv - \mathpzc{K}(a) \mathpzc{V}[\textbf{x}_\mathrm{LPT}](a),
\end{equation}
where the differential operator $\mathpzc{V}[\cdot](a)$ is defined by
\begin{equation}
\mathpzc{V}[\cdot](a) \equiv - \frac{1}{\mathpzc{K}(a)} \frac{\drm}{\drm a} \left( \frac{1}{\mathpzc{D}(a)} \deriv{\,\cdot}{a} \right) .
\end{equation}
With these notations, equation \eqref{eq:kick} reads
\begin{equation}
\deriv{\textbf{p}}{a} = \deriv{\textbf{p}_\mathrm{LPT}}{a} + \deriv{\textbf{p}_\mathrm{res}}{a} = - \mathpzc{K}(a) \mathpzc{V}[\textbf{x}_\mathrm{LPT}](a) + \deriv{\textbf{p}_\mathrm{res}}{a} = \mathpzc{K}(a) \boldsymbol{\nabla} \left( \Delta^{-1} \delta \right) .
\end{equation}
In COLA, the natural variables are therefore $\textbf{x}$ and $\textbf{p}_\mathrm{res}$.

As mentioned in section \ref{ssec:Lagrangian perturbation theory (LPT)}, the key point in COLA is that the fictitious LPT force acting on particles, $\mathpzc{V}[\textbf{x}_\mathrm{LPT}](a)$, can be computed analytically. From equation \eqref{eq:LPT-mapping}, it is straightforward to check that $\mathpzc{V}[\textbf{x}_\mathrm{LPT}](a) = - \mathpzc{V}[D_1](a) \boldsymbol{\Psi}_1 + \mathpzc{V}[D_2](a) \boldsymbol{\Psi}_2$. The computation of $\mathpzc{V}[D_1](a)$ and $\mathpzc{V}[D_2](a)$ uses the differential equations followed by the linear and second-order growth factor, as well as the second Friedmann equation. The result is \citep[see e.g.][equations (1.7), (1.96), (1.118) and appendix B]{LeclercqThesis}
\begin{eqnarray}
\mathpzc{V}[D_1](a) & = & D_1(a) ,\label{eq:VD1} \\
\mathpzc{V}[D_2](a) & = & D_2(a) - D_1^2(a) .\label{eq:VD2}
\end{eqnarray}

The equations of motion to solve are therefore, in tCOLA,
\begin{eqnarray}
\deriv{\textbf{x}}{a} & = & \mathpzc{D}(a)\textbf{p}_\mathrm{res} - \deriv{D_1}{a} \boldsymbol{\Psi}_1 + \deriv{D_2}{a} \boldsymbol{\Psi}_2, \label{eq:drift-cola}\\
\deriv{\textbf{p}_\mathrm{res}}{a} & = & \mathpzc{K}(a) \left[ \boldsymbol{\nabla} \left( \Delta^{-1} \delta \right) - D_1(a) \boldsymbol{\Psi}_1 + \left(D_2(a) - D_1^2(a)\right) \boldsymbol{\Psi}_2 \right]. \label{eq:kick-tcola}
\end{eqnarray}
These are mathematically equivalent to the equations of motion of a PM code (equations \eqref{eq:drift} and \eqref{eq:kick}). In sCOLA, the ``kick equation'' (equation \eqref{eq:kick-tcola}) is replaced for each particle of the sCOLA box by the approximation \citep{Tassev2015}
\begin{equation}
\deriv{\textbf{p}_\mathrm{res}}{a} = \mathpzc{K}(a) \left[ \boldsymbol{\nabla}^\mathrm{sCOLA} \left( (\Delta^\mathrm{sCOLA})^{-1} \delta^\mathrm{sCOLA} \right) - D_1(a) \boldsymbol{\Psi}_1^\mathrm{sCOLA} + \left(D_2(a) - D_1^2(a)\right) \boldsymbol{\Psi}_2^\mathrm{sCOLA} \right]. \label{eq:kick-scola}
\end{equation}
with the notations introduced in section \ref{ssec:Spatial comoving Lagrangian acceleration (sCOLA)}, as well as $\boldsymbol{\Psi}_1^\mathrm{sCOLA} \equiv \boldsymbol{\nabla}_\textbf{q}^\mathrm{sCOLA} \phi^{(1)}$ and $\boldsymbol{\Psi}_2^\mathrm{sCOLA} \equiv \boldsymbol{\nabla}_\textbf{q}^\mathrm{sCOLA} \phi^{(2)}$. Importantly, the ``drift equation'' (equation \eqref{eq:drift-cola}) is not modified, since we are always, by definition, computing a residual displacement with respect to the LPT observer of the full box, whose position is given by equation \eqref{eq:LPT-mapping}.

\section{Standard and modified time-stepping}
\label{apx:Standard and modified time-stepping}

\subsection{Time-stepping in the standard PM algorithm}

In this paper, we adopt the second-order symplectic “kick-drift-kick” algorithm, also known as the leapfrog scheme \citep[e.g][]{Birdsall1985} to integrate the equations of motion. This algorithm relies on integrating the equations on a small time-step and approximating the momenta ($\textbf{p}$ in the ``drift equation'' \eqref{eq:drift}) and accelerations ($\boldsymbol{\nabla}(\Delta^{-1} \delta)$ in the ``kick equation'' \eqref{eq:kick}) that appear in the integrands by their value at some time within the interval. This defines the Drift (D) and Kick (K) operators, which read using the standard discretisation \citep{Quinn1997}:
\begin{eqnarray}
\mathrm{D}(t_i^\mathrm{D},t_f^\mathrm{D},t^\mathrm{K}): & \quad & \textbf{x}(t_i^\mathrm{D}) \mapsto \textbf{x}(t_f^\mathrm{D}) = \textbf{x}(t_i^\mathrm{D}) + \alpha_{\textbf{p}}(t_i^\mathrm{D}, t_f^\mathrm{D}, t^\mathrm{K}) \textbf{p}\!\left(t^\mathrm{K}\right),\label{eq:D_standard}\\
\mathrm{K}(t_i^\mathrm{K},t_f^\mathrm{K},t^\mathrm{D}): & \quad & \textbf{p}(t_i^\mathrm{D}) \mapsto \textbf{p}(t_f^\mathrm{D}) = \textbf{p}(t_i^\mathrm{D}) + \beta_{\delta}(t_i^\mathrm{K}, t_f^\mathrm{K}, t^\mathrm{D}) \left[\boldsymbol{\nabla}\left(\Delta^{-1}\delta\right)\right]\!\!(t^\mathrm{D}),\label{eq:K_standard}
\end{eqnarray}
where
\begin{equation}
\alpha_{\textbf{p}}(t_i^\mathrm{D}, t_f^\mathrm{D}, t^\mathrm{K}) \equiv \int_{t_i^\mathrm{D}}^{t_f^\mathrm{D}} \mathpzc{D}(\tilde{t}) \, \drm \tilde{t} = \int_{t_i^\mathrm{D}}^{t_f^\mathrm{D}} \frac{\drm \tilde{t}}{\tilde{t}^2 \mathcal{H}(\tilde{t})} , \quad \beta_{\delta}(t_i^\mathrm{K}, t_f^\mathrm{K}, t^\mathrm{D}) \equiv \int_{t_i^\mathrm{K}}^{t_f^\mathrm{K}} \mathpzc{K}(\tilde{t}) \, \drm \tilde{t} = -\frac{3}{2} \Omega_\mathrm{m}^{(0)} \int_{t_i^\mathrm{K}}^{t_f^\mathrm{K}} \frac{\drm \tilde{t}}{\tilde{t} \mathcal{H}(\tilde{t})} ,
\label{eq:alpha-beta-standard}
\end{equation}
and $t$ is a function of the scale factor $a$ (typically $t(a)=a$ or $t(a)=\exp(a)$ for time-steps linearly spaced or logarithmically spaced in the scale factor, respectively).

The time evolution between $t_0=t(a_\mathrm{i})$ and $t_{n+1}=t(a_\mathrm{f})$ is then achieved by applying the following operator, $\mathrm{E}(t_{n+1}, t_0)$, to the initial state $(\textbf{x}(t_0), \textbf{p}(t_0))$:
\begin{equation}
\mathrm{K}(t_{n+1/2},t_{n+1},t_{n+1}) \mathrm{D}(t_n,t_{n+1},t_{n+1/2}) \left[ \prod_{i=0}^n \mathrm{K}(t_{i+1/2},t_{i+3/2},t_{i+1}) \mathrm{D}(t_i,t_{i+1},t_{i+1/2}) \right] \mathrm{K}(t_0,t_{1/2},t_0) .
\label{eq:operator_E}
\end{equation}

\subsection{Time-stepping with COLA, standard discretisation}

Using the standard discretisation \citep{Quinn1997} of equations \eqref{eq:drift-cola} and \eqref{eq:kick-tcola}, the Kick and Drift operators for tCOLA are defined by
\begin{eqnarray}
\label{eq:D_COLA}
\mathrm{\widetilde{D}}(t_i^\mathrm{D},t_f^\mathrm{D},t^\mathrm{K}) : & \quad & \textbf{x}(t_i^\mathrm{D}) \mapsto \textbf{x}(t_f^\mathrm{D}) = \textbf{x}(t_i^\mathrm{D}) + \alpha_{\textbf{p}}(t_i^\mathrm{D}, t_f^\mathrm{D}, t^\mathrm{K}) \textbf{p}_\mathrm{res}\!\left(t^\mathrm{K}\right) - \left[ D_1 \right]_{t_i^\mathrm{D}}^{t_f^\mathrm{D}} \boldsymbol{\Psi}_1 + \left[ D_2 \right]_{t_i^\mathrm{D}}^{t_f^\mathrm{D}} \boldsymbol{\Psi}_2, \\
\mathrm{\widetilde{K}}(t_i^\mathrm{K},t_f^\mathrm{K},t^\mathrm{D}) : & \quad & \textbf{p}_\mathrm{res}(t_i^\mathrm{D}) \mapsto \textbf{p}_\mathrm{res}(t_f^\mathrm{D}) = \textbf{p}_\mathrm{res}(t_i^\mathrm{D}) + \beta_{\delta}(t_i^\mathrm{K}, t_f^\mathrm{K}, t^\mathrm{D}) \times \nonumber \\
\label{eq:K_COLA}
& & \hfill \quad\quad\quad\quad\quad\quad\quad \left( \left[ \boldsymbol{\nabla}\left(\Delta^{-1}\delta\right)\right]\!\!(t^\mathrm{D}) - D_1(t^\mathrm{D}) \boldsymbol{\Psi}_1 + \left( D_2(t^\mathrm{D}) - D_1^2(t^\mathrm{D}) \right) \boldsymbol{\Psi}_2 \right) ,
\end{eqnarray}
where the time factors $\alpha_{\textbf{p}}(t_i^\mathrm{D}, t_f^\mathrm{D}, t^\mathrm{K})$ and $\beta_{\delta}(t_i^\mathrm{K}, t_f^\mathrm{K}, t^\mathrm{D})$ are the same as in the PM case (see equation \eqref{eq:alpha-beta-standard}). For sCOLA, $\mathrm{\widetilde{K}}$ is given by the same expression (equation \eqref{eq:K_COLA}) but operates on quantities and differential operators superscripted ``sCOLA'' consistently with equation \eqref{eq:kick-scola}.

In the initial conditions, generated with LPT, we have $\textbf{p} = \textbf{p}_\mathrm{LPT}$; therefore the momentum residual in the rest frame of LPT observers, $\textbf{p}_\mathrm{res}$, should be initialised to zero. At the end, the LPT momentum $\textbf{p}_\mathrm{LPT}$ has to be added to $\textbf{p}_\mathrm{res}$ to recover the full momentum of particles, $\textbf{p}$. This corresponds respectively to the $\mathrm{L}_-$ and $\mathrm{L}_+$ operators \citep[][appendix A]{Tassev2013}, given by
\begin{equation}
\mathrm{L}_\pm(t) : \quad \textbf{p}(t) \mapsto \textbf{p}(t) \pm \textbf{p}_\mathrm{LPT}(t) = \textbf{p}(t) \pm \frac{1}{\mathpzc{D}(t)} \left( -\deriv{D_1}{t} \boldsymbol{\Psi}_1 + \deriv{D_2}{t} \boldsymbol{\Psi}_2 \right) .
\end{equation}

In COLA, the time evolution between $t_0=t(a_\mathrm{i})$ and $t_{n+1}=t(a_\mathrm{f})$ is therefore achieved by applying the following operator to the initial state $(\textbf{x}(t_0),\textbf{p}(t_0))$:
\begin{equation}
\mathrm{L}_+(t_{n+1}) \widetilde{\mathrm{E}}(t_{n+1},t_0) \mathrm{L}_-(t_0),
\end{equation}
where $\widetilde{\mathrm{E}}(t_{n+1},t_0)$ is the operator given by equation \eqref{eq:operator_E}, replacing $\mathrm{D}$ by $\mathrm{\widetilde{D}}$ and $\mathrm{K}$ by $\mathrm{\widetilde{K}}$.

\subsection{Time-stepping with COLA, modified discretisation}

Another approach for the discretisation of equations \eqref{eq:drift-cola} and \eqref{eq:kick-tcola} is proposed by \citet{Tassev2013}. For any arbitrary positive function $u$ of $t$, we can rewrite
\begin{eqnarray}
\deriv{\textbf{x}}{t} & = & \mathpzc{D}(t)u(t) \left\lbrace \frac{1}{u(t)} \times \textbf{p}_\mathrm{res} \right\rbrace - \deriv{D_1}{t} \boldsymbol{\Psi}_1 + \deriv{D_2}{t} \boldsymbol{\Psi}_2,\\
\deriv{\textbf{p}_\mathrm{res}}{t} & = & \deriv{u(t)}{t} \left\lbrace \frac{\mathpzc{K}(t)}{\drm u(t)/\drm t} \times \left[ \boldsymbol{\nabla} \left( \Delta^{-1} \delta \right) - D_1(t) \boldsymbol{\Psi}_1 + \left(D_2(t) - D_1^2(t)\right) \boldsymbol{\Psi}_2 \right] \right\rbrace.
\end{eqnarray}
This form is particularly relevant if $\textbf{p}_\mathrm{res}$ has a time dependence which is entirely captured by a particular $u(t)$, which is universal for all particles. For each equation, considering that the part between curly brackets is constant during the time-step (instead of the momentum and accelerations, respectively), the modified $\mathrm{\widetilde{D}}$ and $\mathrm{\widetilde{K}}$ operators are given by equations \eqref{eq:D_COLA} and \eqref{eq:K_COLA} with the following modified time factors instead of $\alpha_{\textbf{p}}(t_i^\mathrm{D}, t_f^\mathrm{D}, t^\mathrm{K})$ and $\beta_{\delta}(t_i^\mathrm{K}, t_f^\mathrm{K}, t^\mathrm{D})$:
\begin{equation}
\tilde{\alpha}_{\textbf{p}}(t_i^\mathrm{D}, t_f^\mathrm{D}, t^\mathrm{K}) \equiv \frac{1}{u(t^\mathrm{K})} \int_{t_i^\mathrm{D}}^{t_f^\mathrm{D}} \mathpzc{D}(\tilde{t}) u(\tilde{t}) \, \drm \tilde{t}, \quad \tilde{\beta}_{\delta}(t_i^\mathrm{K}, t_f^\mathrm{K}, t^\mathrm{D}) \equiv \left( u(t_f^\mathrm{K}) - u(t_i^\mathrm{K}) \right) \times \frac{\mathpzc{K}(t^\mathrm{D})}{(\drm u(t)/\drm t) (t^\mathrm{D})},
\end{equation}
where in $\tilde{\beta}_{\delta}(t_i^\mathrm{K}, t_f^\mathrm{K}, t^\mathrm{D})$, we have used the trivial integration $\displaystyle{\int_{t_i^\mathrm{K}}^{t_f^\mathrm{K}} \deriv{u(\tilde{t})}{\tilde{t}} \, \drm \tilde{t} = u(t_f^\mathrm{K}) - u(t_i^\mathrm{K})}$.

Using the Ansatz suggested by \citet{Tassev2013}, $u(a)=a^{n_\mathrm{LPT}}$ when $t(a)=a$, we get the explicit expressions
\begin{equation}
\tilde{\alpha}_\textbf{p}(a_i^\mathrm{D}, a_f^\mathrm{D}, a^\mathrm{K}) = \frac{1}{(a^\mathrm{K})^{n_\mathrm{LPT}}} \int_{a_i^\mathrm{D}}^{a_f^\mathrm{D}} \frac{\tilde{a}^{n_\mathrm{LPT}-2}}{\mathcal{H}(\tilde{a})} \, \drm \tilde{a}, \quad \tilde{\beta}_{\delta}(a_i^\mathrm{K}, a_f^\mathrm{K}, a^\mathrm{D}) = -\frac{3}{2} \Omega_\mathrm{m}^{(0)} \frac{(a_f^\mathrm{K})^{n_\mathrm{LPT}} - (a_i^\mathrm{K})^{n_\mathrm{LPT}}}{n_\mathrm{LPT} (a^\mathrm{D})^{n_\mathrm{LPT}} \mathcal{H}(a^\mathrm{D})} .
\end{equation}
We adopt this form and $n_\mathrm{LPT}=-2.5$ for both tCOLA and sCOLA operators, throughout this paper.

\section{Poisson solver with Dirichlet boundary conditions}
\label{apx:Poisson solver with Dirichlet boundary conditions}

In this appendix, we describe how to compute the interior gravitational potential $\Phi$ with Dirichlet boundary conditions. The method is standard in computational physics and has been used at least since \citet{James1977}. Formally, we seek to solve the discrete Poisson equation,
\begin{equation}
\Delta \Phi = \delta,
\label{eq:discretized_Poisson}
\end{equation}
subject to a known boundary potential $\Phi_\mathrm{BCs}$, where $\Delta$ is the FDA to the exact Laplacian operator, i.e. $\Delta \equiv \partial_x^2 + \partial_y^2 + \partial_z^2$ where $\partial_x^2$, $\partial_y^2$, and $\partial_z^2$ are discrete one-dimensional second-order derivatives \citep[see table 1 in][for their expressions in FDA at order 2, 4 and 6]{Hahn2011}.

\subsection{Modified density distribution}

We define $\Phi_\mathrm{BCs}$ as having non-zero values only in a layer of $N_\mathrm{ghost}$ cells immediately outside the active domain of the PM grid. We can then write the desired potential as $\Phi \equiv \widetilde{\Phi} + \Phi_\mathrm{BCs}$ where the required boundary condition for $\widetilde{\Phi}$ is $\widetilde{\Phi} = 0$. From the definition of $\Phi_\mathrm{BCs}$, $\Delta \Phi_\mathrm{BCs}$ will be non-zero only in a layer of $N_\mathrm{ghost}$ active cells just inside the domain boundaries. We can thus define a modified density distribution,
\begin{equation}
\delta' \equiv \delta - \Delta \Phi_\mathrm{BCs},
\end{equation}
which is the same as $\delta$ everywhere except in the layer of $N_\mathrm{ghost}$ cells adjoining the domain boundaries. We can then employ a zero-boundary condition Poisson solver to obtain a solution of $\Delta \widetilde{\Phi} = \delta'$ (see section \ref{sapx:Zero-boundary condition Poisson solver}). Within the interior, where $\Phi_\mathrm{BCs} = 0$, this solution is the desired final solution of $\Delta \Phi = \delta$ with the Dirichlet boundary condition $\Phi = \Phi_\mathrm{BCs}$.

\subsection{Zero-boundary condition Poisson solver} 
\label{sapx:Zero-boundary condition Poisson solver} 

In cosmological simulations, it is conventional to use FFTs to solve the Poisson equation, since the discrete Laplacian operator is diagonal in Fourier space. FFTs assume that the input source has periodic boundary conditions. Similarly, for zero boundary conditions, we can work with three-dimensional type-I discrete sine transforms (DST-I), defined by
\begin{equation}
\delta^{\ell,m,n} \equiv \sum_{i=1}^{N_x}\sum_{j=1}^{N_y}\sum_{k=1}^{N_z} \delta_{i,j,k} \mathcal{X}_i^\ell \mathcal{Y}_j^m \mathcal{Z}_k^n,
\label{eq:DST_delta}
\end{equation}
where $\delta_{i,j,k}$ is the value of the source field in the voxel indexed by $1 \leq i \leq N_x$, $1 \leq y \leq N_y$, $1 \leq k \leq N_z$ ($N_x=N_y=N_z=N_\mathrm{g}$ in this paper). The basis functions are defined by
\begin{equation}
\mathcal{X}_i^\ell \equiv \sin\left( \frac{\pi \ell i}{N_x+1} \right), \quad \mathcal{Y}_j^m \equiv \sin\left( \frac{\pi m j}{N_y+1} \right), \quad \mathcal{Z}_k^n \equiv \sin\left( \frac{\pi n k}{N_z+1} \right) .
\end{equation}
They ensure that the signal has zero boundary values (for $i \in \{0,N_x+1\}$ or $j \in \{0,N_y+1\}$ or $k \in \{0,N_z+1\}$). They satisfy discrete orthogonality relations, for example,
\begin{equation}
\frac{2}{N_x+1} \sum_{i=1}^{N_x} \mathcal{X}_i^\ell \mathcal{X}_i^{\ell'} = \updelta_\mathrm{K}^{\ell \ell'} \quad \mathrm{and} \quad \frac{2}{N_y+1} \sum_{m=1}^{N_y} \mathcal{Y}_j^m \mathcal{Y}_{j'}^m = \updelta_\mathrm{K}^{j j'},
\end{equation}
where $\updelta_\mathrm{K}$ is the Kronecker symbol. The inverse transformation is simply DST-I multiplied by $8/\left[ (N_x+1)(N_y+1)(N_z+1) \right]$, i.e.
\begin{equation}
\delta_{i,j,k} = \frac{8}{(N_x+1)(N_y+1)(N_z+1)} \sum_{i=1}^{N_x}\sum_{j=1}^{N_y}\sum_{k=1}^{N_z} \delta^{\ell,m,n} \mathcal{X}_i^\ell \mathcal{Y}_j^m \mathcal{Z}_k^n ,
\label{eq:IDST_delta}
\end{equation}
and similarly for the gravitational potential,
\begin{equation}
\Phi_{i,j,k} = \frac{8}{(N_x+1)(N_y+1)(N_z+1)} \sum_{i=1}^{N_x}\sum_{j=1}^{N_y}\sum_{k=1}^{N_z} \Phi^{\ell,m,n} \mathcal{X}_i^\ell \mathcal{Y}_j^m \mathcal{Z}_k^n .
\label{eq:IDST_Phi}
\end{equation}

It is straightforward to show that $\mathcal{X}_i^\ell$, $\mathcal{Y}_j^m$, $\mathcal{Z}_k^n$ are eigenfunctions of the discrete one-dimensional second-order derivatives $\partial_x^2$, $\partial_y^2$, and $\partial_z^2$, respectively. The corresponding eigenvalues $\lambda_x^\ell$, $\lambda_y^m$ and $\lambda_z^n$ are given by
\begin{eqnarray}
\lambda_x^\ell & = & -\frac{4}{d_x^2} \sin^2\left( \frac{k_\ell \, d_x}{4} \frac{N_x}{N_x+1} \right), \\
\lambda_x^\ell & = & \frac{1}{3 d_x^2} \left[\sin^2\left( \frac{k_\ell \, d_x}{2} \frac{N_x}{N_x+1} \right) - 16\sin^2\left( \frac{k_\ell \, d_x}{4} \frac{N_x}{N_x+1} \right) \right], \\
\lambda_x^\ell & = & -\frac{1}{45d_x^2} \left[2\sin^2\left( \frac{3k_\ell \, d_x}{4} \frac{N_x}{N_x+1} \right) - 27\sin^2\left( \frac{k_\ell \, d_x}{2} \frac{N_x}{N_x+1} \right) + 270\sin^2\left( \frac{k_\ell \, d_x}{4} \frac{N_x}{N_x+1} \right) \right],
\end{eqnarray}
for FDA at order 2, 4, and 6 respectively, where $k_\ell \equiv 2\pi \ell/L_x$, $d_x \equiv L_x/N_x$ and $L_x$ is the size of the box along the $x$-direction ($L_x = L_\mathrm{sCOLA} \equiv L/N_\mathrm{tiles}$ in this paper). Similar expressions exist for $\lambda_y^m$ and $\lambda_z^n$.

Plugging equations \eqref{eq:IDST_delta} and \eqref{eq:IDST_Phi} into \eqref{eq:discretized_Poisson} and using the orthogonality relations, we obtain a simple form for the discretised Poisson equation in sine space,
\begin{equation}
\Phi^{\ell,m,n} = \frac{\delta^{\ell,m,n} }{\lambda_x^\ell + \lambda_y^m + \lambda_z^n}.
\label{eq:Poisson_sine_space}
\end{equation}
Therefore, the Poisson equation $\Delta \Phi = \delta$ with zero boundary conditions can be solved by the following three steps:
\begin{enumerate}
\item performing a forward DST of the source ($\delta_{i,j,k} \rightarrow \delta^{\ell,m,n}$), according to equation \eqref{eq:DST_delta} (costing $\mathcal{O}(N_x N_y N_z \log\left[ N_x N_y N_z \right])$ operations),
\item solving the Poisson equation in sine space ($\delta^{\ell,m,n} \rightarrow \Phi^{\ell,m,n}$), according to equation \eqref{eq:Poisson_sine_space} (costing $\mathcal{O}(N_x N_y N_z)$ operations),
\item performing an inverse DST of the gravitational potential ($\Phi^{\ell,m,n} \rightarrow \Phi_{i,j,k}$), according to equation \eqref{eq:IDST_Phi} (costing $\mathcal{O}(N_x N_y N_z \log\left[ N_x N_y N_z \right])$ operations).
\end{enumerate}
In practice, forward and inverse DSTs are performed using the FFTW library \citep{FFTW05}, publicly available at \url{http://www.fftw.org}, where the DST-I is known as $\texttt{FFTW\_RODFT00}$.

\vspace{25pt}
\twocolumngrid

\section*{ORCID iDs}

\noindent Florent Leclercq \orcid{0000-0002-9339-1404} \href{https://orcid.org/0000-0002-9339-1404}{0000-0002-9339-1404}\newline
\noindent Baptiste~Faure \orcid{0000-0002-6623-2344} \href{https://orcid.org/0000-0002-6623-2344}{0000-0002-6623-2344}\newline
\noindent Guilhem Lavaux \orcid{0000-0003-0143-8891} \href{https://orcid.org/0000-0003-0143-8891}{0000-0003-0143-8891}\newline
\noindent Benjamin~D.~Wandelt \orcid{0000-0002-5854-8269} \href{https://orcid.org/0000-0002-5854-8269}{0000-0002-5854-8269}\newline
\noindent Andrew~H.~Jaffe \orcid{0000-0003-2086-1759}
\href{https://orcid.org/0000-0003-2086-1759}{0000-0003-2086-1759}\newline
\noindent Alan~F.~Heavens \orcid{0000-0003-1586-2773} \href{https://orcid.org/0000-0003-1586-2773}{0000-0003-1586-2773}\newline
\noindent Will~J.~Percival \orcid{0000-0002-0644-5727} \href{https://orcid.org/0000-0002-0644-5727}{0000-0002-0644-5727}\newline
\noindent Camille~Noûs \orcid{0000-0002-0778-8115} \href{https://orcid.org/0000-0002-0778-8115}{0000-0002-0778-8115}

\section*{Statement of contribution}

FL and BDW conceived the project. FL wrote the \textsc{Simbelmynë} code, implemented the new sCOLA algorithm, ran the simulations, performed the study, supervised BF's internship project, and wrote the bulk of the paper. BF contributed to the first implementation of the tiling algorithm in \textsc{Simbelmynë} and to early tests of the method. GL suggested the investigation of an alternative Poisson solver, proposed tests of the impact of boundary effects, and contributed to writing the paper. BDW made conceptual contributions, helped designing the accuracy and speed tests, and contributed to writing the paper. AHJ prompted the use of the linearly-evolving potential approximation. AHJ and AFH contributed to the interpretation of results. WJP supported the design of the first version of the algorithm and contributed to student supervision. CN contributed to the collegial construction of the standards of science, by developing the methodological framework, the state-of-the-art, as well as post-publication procedures. All natural authors read and approved the final manuscript.

\section*{Acknowledgements}

We are grateful to Mat\'ias Zaldarriaga for stimulating discussions and useful comments throughout the realisation of this project. We thank Jens Jasche and Svetlin Tassev for discussions that triggered this project, and Oliver Hahn for constructive observations. FL and BDW acknowledge the hospitality of the Institute for Advanced Study, Princeton, where this project was initiated. FL, BF and WJP thank the Institute of Cosmology and Gravitation of the University of Portsmouth, where part of this work was prepared.

This work made use of NumPy \citep{NumPy}, IPython \citep{IPython}, Matplotlib \citep{Matplotlib}, Jupyter notebooks \citep{Jupyter}, and the colourmaps provided by the cmocean \citep{cmocean} and CMasher (\url{https://github.com/1313e/CMasher}) packages.

FL acknowledges funding from the Imperial College London Research Fellowship Scheme. GL and BDW acknowledge financial support from the ANR BIG4, under reference ANR-16-CE23-0002. The Center for Computational Astrophysics is supported by the Simons Foundation. Research at Perimeter Institute is supported in part by the Government of Canada through the Department of Innovation, Science and Economic Development Canada and by the Province of Ontario through the Ministry of Colleges and Universities. Numerical computations were done on the Sciama High Performance Compute (HPC) cluster which is supported by the ICG, SEPNet and the University of Portsmouth; and on the cx1 cluster hosted by the Research Computing Service facilities at Imperial College London (\href{http://doi.org/10.14469/hpc/2232}{doi:10.14469/hpc/2232}). This work was done within the Aquila Consortium (\url{https://aquila-consortium.org}).

\section*{References}
\bibliography{biblio}

\end{document}